\documentclass[12pt,preprint]{aastex}
\usepackage{psfig,lscape}
\def\lap{\lower.5ex\hbox{$\; \buildrel < \over \sim \;$}}
\def\gap{\lower.5ex\hbox{$\; \buildrel > \over \sim \;$}}

\def\ergcm2s{${\rm erg\ cm^{-2}\ s^{-1}}$}
\def\cs{${\rm s^{-1}}$}
\def\ergscm2s{${\rm erg\ cm^{-2}\  s^{-1}}$}

\def\cm-2{${\rm cm^{-2}}$}
\def\ergs{${\rm erg\ s^{-1}}$}

\def\sax1808{SAX~J1808.4-3658}
\def\msol   {{M$_{\odot}$}}

\def\t3 {r2-67}

\received{2003 June 20}
\begin{document}
\title{A Synoptic X-ray Study of M31 with the {\it Chandra}-HRC}
\author{Benjamin F. Williams\altaffilmark{1}, Michael R. Garcia \altaffilmark{1}, Albert K. H. Kong\altaffilmark{1}, Frank A. Primini\altaffilmark{1}, A. R. King\altaffilmark{2}, Rosanne Di Stefano\altaffilmark{1}, and Stephen S. Murray\altaffilmark{1}}
\altaffiltext{1}{Harvard-Smithsonian Center for Astrophysics, 60
Garden Street, Cambridge, MA  02138; williams@head-cfa.harvard.edu;
garcia@head-cfa.harvard.edu; akong@head-cfa.harvard.edu; fap@head-cfa.harvard.edu; rd@head-cfa.harvard.edu; ssm@head-cfa.harvard.edu}
\altaffiltext{2}{Theoretical Astrophysics Group, University of Leicester, University Road, Leicester LE1 7RH, UK; ark@astro.le.ac.uk}

\begin{abstract}

We have obtained 17 epochs of {\it Chandra} High Resolution Camera
(HRC) snapshot images, each covering most of the M31 disk.  The data
cover a total baseline of $\sim$2.5 years and contain a mean effective
exposure of 17 ks.  We have measured the mean fluxes and long-term
lightcurves for 166 objects detected in these data.  At least 25\% of
the sources show significant variability.  The cumulative luminosity
function (CLF) of the disk sources is well-fit by a power-law with a
slope comparable to those observed in typical {\it elliptical}
galaxies.  The CLF of the bulge is a broken power law similar to
measurements made by previous surveys.  We note several sources in the
southwestern disk with ${\rm L_X} > 10^{37}$\ergs .  We
cross-correlate all of our sources with published optical and radio
catalogs, as well as new optical data, finding counterpart candidates
for 55 sources.  In addition, 17 sources are likely X-ray transients.
We analyze follow-up HST WFPC2 data of two X-ray transients, finding
F336W ($U$-band equivalent) counterparts.  In both cases, the
counterparts are variable.  In one case, the optical counterpart is
transient with F336W $= 22.3\pm 0.1$ mag.  The X-ray and optical
properties of this object are consistent with a $\sim$10 solar mass
black hole X-ray nova with an orbital period of 23$^{+54}_{-16}$ days.
In the other case, the optical counterpart varies between F336W =
20.82$\pm$0.06 mag and F336W = 21.11$\pm$0.02 mag.  Ground-based and
HST observations show this object is bright ($V$ = 18.8$\pm$0.1) and
slightly extended.  Finally, the frequency of bright X-ray transients
in the M31 bulge suggests that the ratio of neutron star to black hole
primaries in low-mass X-ray binaries ($NS/BH$) is $\sim$1.

\end{abstract}

\section{Introduction}

M31 contains hundreds of X-ray sources in a relatively small field.
Precision measurement of their positions allows identification of
optical and radio counterparts.  Long-term monitoring of these sources
provides variability information on timescales which are not probed by
single observations.  This information can help determine the nature
of the X-ray sources.  For example, X-ray binaries containing high and
low mass secondaries have somewhat different variability, and SNR are
not expected to show any variability at all.  By using the X-ray
variability and luminosity information to determine the nature of the
sources one can establish links to the stellar populations in which
the sources reside.  These links include the effects of star formation
on the X-ray source population and the effects of galaxy evolution on
X-ray source production.

Several surveys of M31 have been completed in X-rays, finding hundreds
of sources, a large fraction of which are variable.  Most surveys have
concentrated on the central bulge region, which contains most of the
bright X-ray sources.  These studies began with
\citet{vanspeybroeck1979}, who used Einstein data to catalog 69
objects brighter than $\sim$9$\times$10$^{36}$\ergs\/ in the
M31 bulge and northern M31 disk. \citet{collura1990} found 2 variable
X-ray point sources in the Einstein data of M31.  Later
\citet{trinchieri1991} performed a deeper survey of more than half the
M31 disk by combining all Einstein data of M31.  They found 108
sources brighter than $\sim$5$\times$10$^{36}$\ergs\/ including
fourteen additional variable sources.  The central 34 arcmin of M31
was surveyed with the ROSAT HRI \citep{primini1993}, revealing 86
sources brighter than $\sim$10$^{36}$\ergs .  By comparison
with previous Einstein observations, they found nearly half of the
sources in the bulge to be variable.  Two more ROSAT surveys were
completed with the PSPC \citep{supper1997,supper2001}.  These surveys
together covered most of the disk (10.7 deg$^2$) and revealed 560
X-ray sources down to a detection limit of $\sim$5$\times$10$^{35}$
\ergs ; they found 34 sources varied in the 1 year between
observations.  

Recently, M31 has been studied with the {\it Chandra} X-ray
Observatory and XMM-{\it Newton}.  The improved resolution and
sensitivity have led to additional interesting observations.  For
example, {\it Chandra} observations have revealed several new X-ray
transients \citep{atel97,atel82,atel79,atel76,iauc7291}, as have XMM
observations \citep{iauc7798,iauc7659}.  \citet{trudolyubov2001}
discussed XMM-{\it Newton} and {\it Chandra} observations of three of
these which were discovered in the year 2000.  Using XMM-{\it Newton}
observations, \citet{barnard2003} showed that the variability
properties of one of the brighter sources in M31 indicate that it is a
stellar mass black hole binary.  \citet{kong2002} performed a survey
of the central $\sim$17$' \times 17'$ of M31 with the {\it Chandra}
ACIS-I, finding 204 X-ray sources down to a detection limit of
$\sim$2$\times$10$^{35}$\ergs .  About half of the sources were
variable on timescales of months, and 13 sources were transients.
\citet{kaaret2002} used HRC-I data of the nuclear region to detect 142
sources brighter than $\sim$2$\times$10$^{35}$\ergs , finding nearly
half of the bright sources to be variable on timescales of $\lap$10
hours. \citet{trudolyubov2002} performed a deep XMM-{\it Newton}
survey of the northern half in the disk, finding that the M31 disk is
deficient in bright X-ray sources.  Further XMM-{\it Newton}
observations have discovered diffuse soft X-ray emission associated
with the northern disk \citep{trudolyubov2004}.  \citet{kong2002b}
surveyed three widely separated portions of the M31 disk with ACIS-I,
finding possible differences between the X-ray source populations in
these different regions.  \citet{rd.gc.2002} found that M31 globular
clusters can be more X-ray luminous than those of the Galaxy and
suggested that this was due in part to the larger number of M31
globular clusters (GCs) rather than a difference in the shape of the
luminosity function (LF).  Finally, \citet{distefano2003} have
completed a survey for supersoft X-ray sources (SSSs) and quasisoft
sources (QSSs) in four regions of M31, finding 33 such objects.

We have obtained {\it Chandra} HRC-I data covering most of the M31
optical disk. These data provide the first opportunity to perform a
large area survey of M31 with {\it Chandra}, including regular
information about the state of the detected sources over a period of
two and a half years.  While the sensitivity and coverage (0.9
deg$^2$) are not as extensive as that of \citet{supper1997,supper2001}
which covered an area of 10.7~deg$^2$, the time baseline is well
sampled.  We have also obtained 3-epoch HST (WFPC2) images of 2 newly
discovered X-ray transients in order to search for their optical
counterparts.  Among the deepest ground based images of M31 are those
that were obtained as part of the Local Group Survey project
\citep{massey2001}; we analyze unpublished sections of these data in
order to search for new optical counterparts.  In this paper we use
the {\it Chandra} data to create an X-ray source catalog covering most
of M31 and to measure long-term variability in the X-ray emission from
these sources, and we use the newly obtained optical data to search
for long wavelength counterparts.  In \S 2 we present the X-ray data,
source list and lightcurves.  In \S 3, we discuss the X-ray results,
including LFs and variability studies.  In \S 4 we describe the
optical data used to search for counterparts to the X-ray sources.  In
\S 5 we discuss the results of the search for counterparts.  Section 6
describes our detailed analysis of 2 X-ray transient sources detected
in the optical with HST.  Finally, \S 7 provides a summary of our
conclusions.

\section{X-ray Observations}

\subsection{HRC Observations and Data Reduction}
\begin{figure}
\centerline{\psfig{file=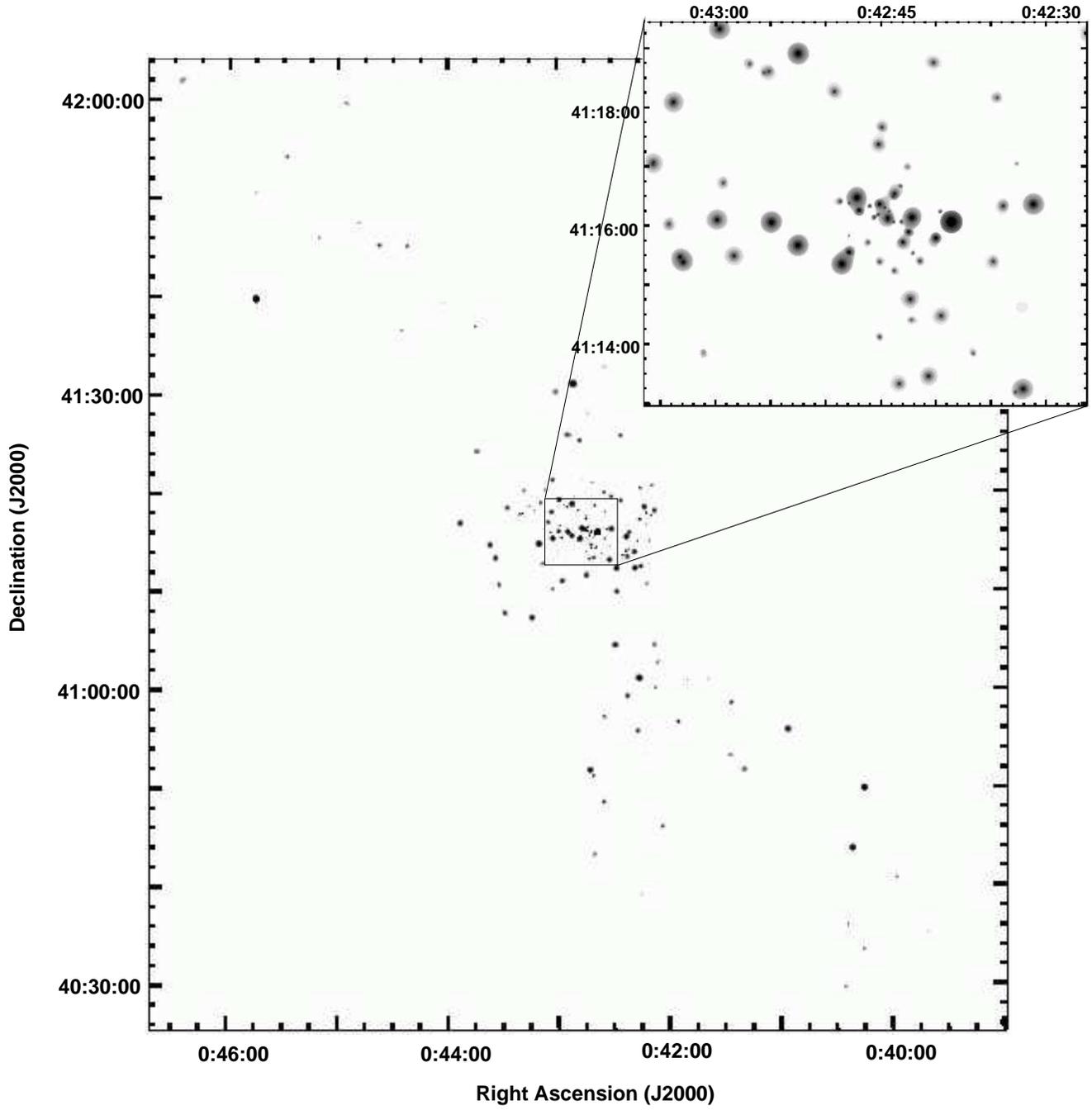,height=7in,angle=0}}
\caption{An exposure corrected, background subtracted source image
containing all 80 {\it Chandra} HRC exposures used for this study is shown.
Axes provide the R.A. and Dec. of the area shown.  The inset image
shows the central bulge region at higher spatial resolution.
}
\label{sim}
\end{figure}

The data for this project are originally part of a survey program to
look for X-ray transients in M31.  Nearly every month from November of
1999 to February of 2001, {\it Chandra} took HRC-I images of 5 fields
covering most of M31.  Observations were then made every few months
until June of 2002.  Each image was shallow ($\sim$1 ks) but
sufficient to detect any strong X-ray transients in the observed
epoch. Individual epochs of these data were of limited use for survey
purposes because of their short exposure times, but herein we stack
the data into a deeper 17 ks HRC-I mosaic of M31. The observation ID
(OBSID) numbers, dates, exposure times and pointing coordinates of all
of the observations are given in Table~\ref{obslist}. An exposure map
of the stacked data is shown in Figure~\ref{expmap}, where the thick
white outline marks the region of the data where the 6$\sigma$
sensitivity is at least 1.3$\times$10$^{37}$\ergs (0.9 deg$^2$). Our
LFs are complete to this luminosity (see \S 3.1).  We focused on this
region in our analysis to provide a constant area of known sensitivity
limit.  The combined, exposure corrected, background subtracted source
image is shown in Figure~\ref{sim}. The image shows the majority of
the sources are near the center of M31.  The areas near center of the
galaxy with the highest exposure have a $3.5\sigma$ detection limit of
$\sim$1.5 $\times$ 10$^{36}$\ergs .

In order to combine the data, they were first aligned using the CIAO
script {\it{align\_evt}}, which corrects for small errors in the
aspect solutions of different exposures by aligning the detected
sources in the field.  The images of the bulge contained $\sim$5
sources suitable for this purpose, which allowed alignment to an rms
accuracy of $0.3''$.  Unfortunately, in the disk, where there are
fewer bright sources, this technique was less successful.  In these
outer regions, the fields were aligned using the {\it Chandra} aspect
solution, which is accurate to $\sim$1$''$.  In the uncrowded outer
regions, we binned the data to a resolution of 2$''$.  This binning
provided better detection of faint objects by removing the effects of
the less precise alignment between exposures in the disk fields.  In
the central 18$'$ $\times$ 18$'$ of the galaxy, where the alignment
was better, we binned the data to 1$''$ resolution in order to better
match the instrumental point spread function (PSF).  Finally, we
combined the data into 3 data sets using the task {\it{merge\_all}}.
One set contained the data for the northern half of the galaxy,
another contained the southern half and the last contained the center.

\subsection{Source List}

\subsubsection{Source Detection}

We searched for sources in the three data sets using the CIAO task
{\it{wavdetect}}.  We ran this task searching for sources on 4 size
scales: 1, 2, 4 and 8 pixels.  The pixels in the merged images were
1$''$ in the central 18$'$ $\times$ 18$'$ and 2$''$ outside of this
region.  By searching on several scales {\it{wavdetect}} is able to
overcome the large changes in the size of of the {\it Chandra} PSF
from about 0.5$''$ near the center of the field to over 10$''$ in the
outer regions of the field.  One hundred sixty-six sources were
detected above our 3.5$\sigma$ detection threshold.  Their short
names, positions, detection counts, signal-to-noise ratios ($S/N$),
mean X-ray luminosities (${\rm L_X}$; 0.1--10 keV), $\chi^2_{\nu}$
values for a fit to a constant source, counterparts or previous X-ray
detections, and references for those counterparts and previous
detections are provided in Table~\ref{slist}.  Detailed descriptions
of all columns, including a definition of the source naming convention
(r1, r2, n1, s2, etc.) are given in the notes to Table~\ref{slist}.
To convert our measured mean X-ray fluxes to luminosities, we assumed
a distance to M31 of 780 kpc, a photon index $\alpha = 1.7$, and
absorption ${\rm N_H} = 10^{21}$ cm$^{-2}$.  These assumptions
provided a constant conversion factor of 2.5 $\times$ 10$^{41}$ erg
cm$^2$ ct$^{-1}$ from HRC flux in ct cm$^{-2}$ s$^{-1}$ to luminosity
(\ergs\/ 0.1--10 keV).

\begin{figure}
\centerline{\psfig{file=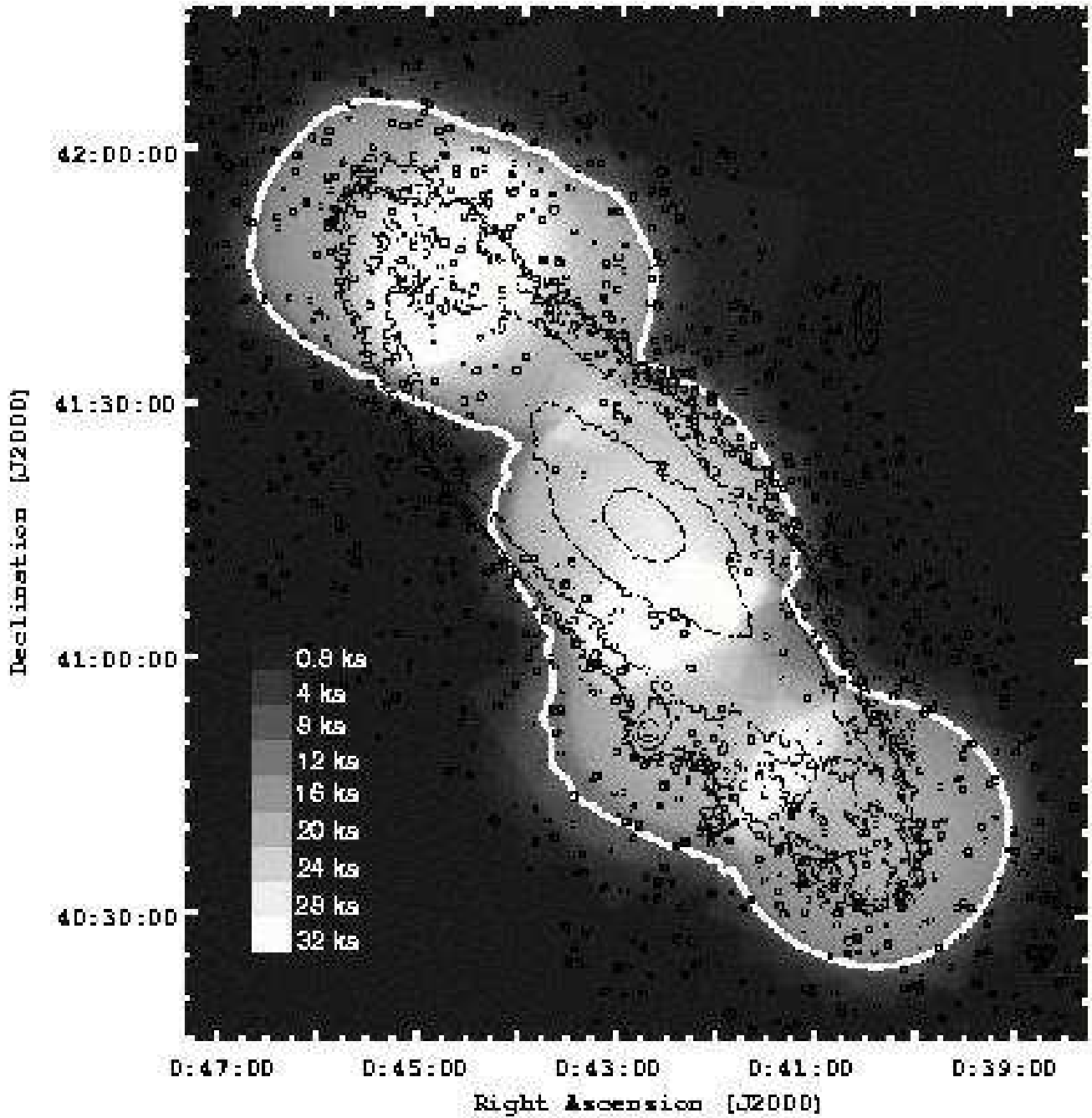,height=7in,angle=0}}
\caption{An exposure map of the 80 {\it Chandra}-HRC exposures used
for this study is shown (greyscale).  Optical contours of the M31 disk
are plotted on the map to show the area covered.  The heavy white
outline marks the region analyzed, where the limiting luminosity does
not exceed 1.3$\times$10$^{37}$ \ergs.  Axes provide the
R.A. and Dec. of the area shown.  Labeled boxes provide a key to the
exposure times represented by the greyscale. }
\label{expmap}
\end{figure}

We consulted previous surveys of the X-ray source background to set a
limit on the number of possible background sources in our sample.
ROSAT observations of the Lockman Hole \citep{hasinger1998} show that
there are $\lap$40 sources per square degree with luminosities greater
than 4$\times$10$^{36}$\ergs .  Our large-area disk LF quickly falls
off below this luminosity due to completeness (cf. \S 3.1).  Therefore
we estimate that our entire catalog contains $<$40 background and/or
foreground sources.  Near the center of M31, our data contain an area
of 0.01 deg$^2$ complete at 6$\sigma$ to $\sim$2.5
$\times$ 10$^{36}$ \ergs (see the area of highest exposure near the
center of M31 in Figure~\ref{expmap}). This area should contain
$\sim$100 background sources per deg$^2$ \citep{hasinger1998}, or 1
source in our most sensitive area, where we detect 41 sources.
Therefore contamination near the galaxy center is very small.

Within the central 10 arcmin of M31, all but one of the non-transient
sources in our list has been previously detected in the X-ray band.
This region has been well studied with {\it Chandra} already
\citep{kong2002,kaaret2002}.  Source r2-68 appears to be a clean new
detection, with 19 counts and a S/N ration of 4.6.  While we do not
detect any variations in the HRC snapshots of this source
($\chi^2_{\nu} = 0.15$), the fact that it is not detected in the much
deeper ACIS \citep{kong2002} and 50ks~HRC exposures \citep{kaaret2002}
indicates it is variable.  As shown in Figure~\ref{varsen} and
discussed in \S 3.2, our sensitivity to variability in the fainter
sources of our survey, such as r2-68, is limited.

Outside of the central region, we detect several new objects.  Of the
new objects, s1-76, s1-82, and n1-84 are well detected, with $S/N$ of
9.3, 8.4, and 4.8 respectively.  These objects are all brighter than
10$^{36}$ \ergs ; therefore their non-detection in the ROSAT surveys
(which reached 5$\times$10$^{35}$~\ergs ) indicates that they could be
variable on timescales as long as a decade.  While we do not detect
variability in these four sources between the individual HRC-I
snapshots, the counting rates are low enough that variations of a
factor of two may be undetected.  In addition, there are several new
sources that varied significantly during our survey.  Among these
variable sources are s1-79, which is a strong transient with a peak
luminosity of 1.5$\times$10$^{38}$~\ergs\ and peak $S/N$ of 35, and
s1-1, which is variable on year long timescales with a $\chi^2$ test
for a steady source giving $\chi^2_{\nu}$ of 1.4 (probability = 13\%).

\subsection{Lightcurves}

By combining the data from our fields, we were able to construct
lightcurves covering more than 2 years for all of the objects we
detected.  Initial lightcurves for the objects were measured using the
CIAO task {\it{lightcurve}}, which measures the number of counts in a
square aperture around the detected object in each epoch. We measured
source fluxes in each of the 17 epochs detailed in column 3 of
Table~\ref{obslist}.  The lightcurves were measured using boxes with
0.7$''$ sides within 3$''$ of the nucleus where the sources are
separated by only 1$''$.  This box size was increased to 2$''$ in the
central 1$'$, where the sources are generally separated by more than
2$''$.  The box size was increased to 8$''$ in the central 18$'$
$\times$ 18$'$ (outside of the central 1$'$), where the sources are
generally separated by more than 8$''$.  Outside of this region, the
sources are generally separated by over an arcminute.  Because these
sources are typically located farther off-axis, where the PSF is
larger, we used a 16$''$ box to measure them.  The average radius of a
circle that encloses 90\% of the energy at 1.49 keV in the HRC-I is
1$''$ at 1$'$ off-axis, 4$''$ at 6$'$ off-axis, and 8$''$ at 8$'$
off-axis.  These box sizes therefore insured that we measured most of
the source counts in off-axis areas where crowding of the sources was
not an issue.  The 8$''$ box size typically contained only $\sim$1
background count per ks timebin, while the 16$''$ boxes typically
contained $\sim$4 background counts per ks timebin.

The background subtracted count rates from {\it{lightcurve}} were used
as a starting point for our lightcurve measurement for each source.
We converted the {\it{lightcurve}} output from units of counts to flux
units using exposure maps which take into account the aspect
histograms and instrumental flat field.  As the instrument's effective
area depends on the source spectrum, we made the exposure maps
assuming a typical M31 source spectrum of an absorbed power law with
an index of 1.7 and an absorption column density of $10^{21}$
cm$^{-2}$.  These maps were created for each epoch of the lightcurve
to measure the effective exposure for each object in each epoch
individually.  The {\it{lightcurve}} output was then converted to flux
units using the effective exposure of each object in each observation.
Finally, the fluxes were multiplied by our conversion factor of 2.5
$\times$ 10$^{41}$ erg cm$^2$ ct$^{-1}$ to estimate the source
luminosity.  Table~\ref{slist} lists the $\chi^2_{\nu}$ fit of each
lightcurve to a constant flux at its mean value.  The objects have a
wide range of variability.  Sources with $\chi^2_{\nu}$ values of
greater than 1.47 are discussed as variables in \S 3.2.  Several of
those with the highest $\chi^2_{\nu}$ values ($\gap$10) are
transients, discussed in detail in \S 3.2 and \S 6.

\begin{figure}
\centerline{\psfig{file=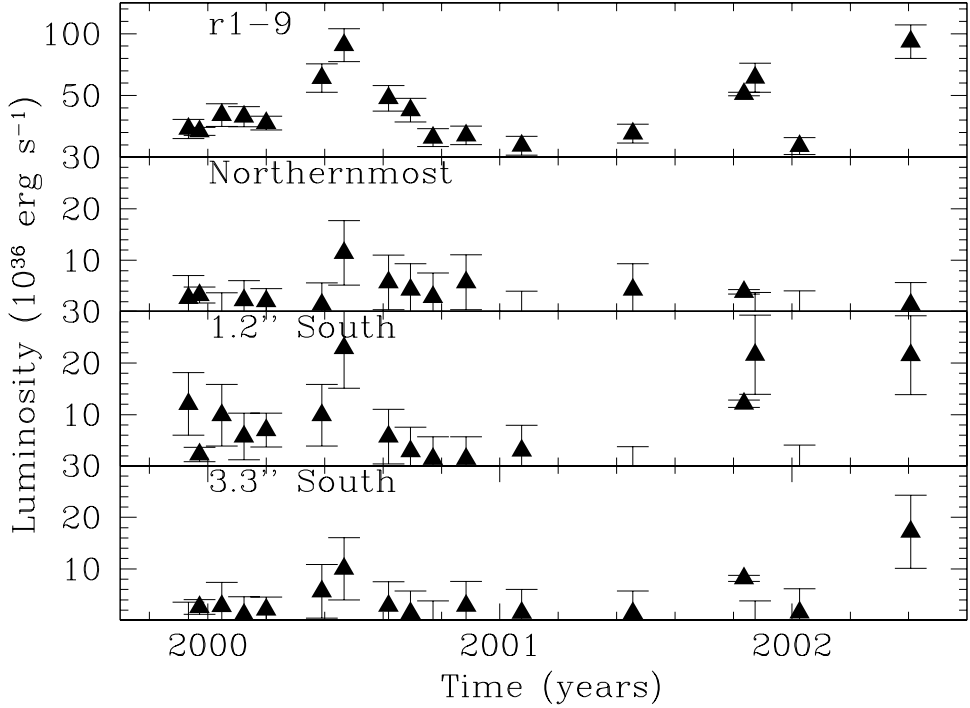,height=4.0in,angle=270}}
\caption{Long term lightcurve of the center of M31 is shown in the top
panel.  This lightcurve uses a large aperture to maximize photon
counts.  It actually contains the light of 3 sources.  The individual
lightcurves, with fewer counts, are shown in the three lower panels.
These lightcurves were measured using boxes with 0.7$''$ sides.  The
second curve from the top shows the northernmost source (closest to
the nucleus), which shows no evidence for variability even on this
long timescale.  The brighter source 1.2$''$ south of the nucleus
(third lightcurve) shows the strong variability seen in the large
aperture lightcurve (previous figure), but the counts are fewer in the
small aperture photometry.  The southernmost source (bottom curve)
shows similar variability, so that these two sources must be
contaminating each other in our data set.}
\label{nucleus}
\end{figure}

The 2$''$ box size used for the central arcmin of M31 was appropriate
for measuring the lightcurves for all of the central sources except
for the well-known 3 bright sources in the nucleus of the galaxy
\citep{garcia2000}.  Our {\it wavdetect} analysis detected these
sources as the single (extended) source r1-9.  The lightcurve of these
three sources blended together is shown in the top panel of
Figure~\ref{nucleus}.  The combined lightcurve reveals variability of
the nuclear region by a factor of $\sim$10 on timescales of about 6
months.  We attempted to create 3 separate lightcurves for the 3 known
objects by dividing the blended nuclear region into 3 parts.  We made
lightcurves for each of these parts using a boxsize of 0.7$''$.  While
the northernmost of these three is closest to the nucleus
\citep{garcia2001}, it is unclear if the source is associated with the
central massive black hole.

The 3 lightcurves are shown in Figure~\ref{nucleus}, but the
lightcurves appear to be influenced by one another, revealing the
spatial limitations of our data set.  Even so, we see that the part
closest to the nucleus, corresponding to CXOM31~J004244.3+411608,
shows the least variability.  It appears as constant and faint
throughout the observations.  The highest level of variability among
the three central sources is shown by the bright, soft source
immediately to the south (CXOM31~J004244.3+411607) which was initially
(and incorrectly) associated with the central black hole
\citep{garcia2000}.  The lightcurve of the source furthest to the
south (CXOM31 J004244.3+411605) shows less variability but mirrors
that of its neighbor to the north, suggesting that the two are not
completely resolved.  In addition to looking for long-term variability
in the nuclear region, we were also able to look for short-term
variability using the long exposure of \citet{kaaret2002}, where 3
sources are clearly resolved.  The lightcurves of the three objects
showed no variability on this timescale.  They all had $\chi^2_{\nu}$
values of less than 1 when fit to a constant flux, including the soft
source which is so highly variable on longer timescales
($\chi^2_{\nu}$ = 18.4).

\begin{figure}
\centerline{\psfig{file=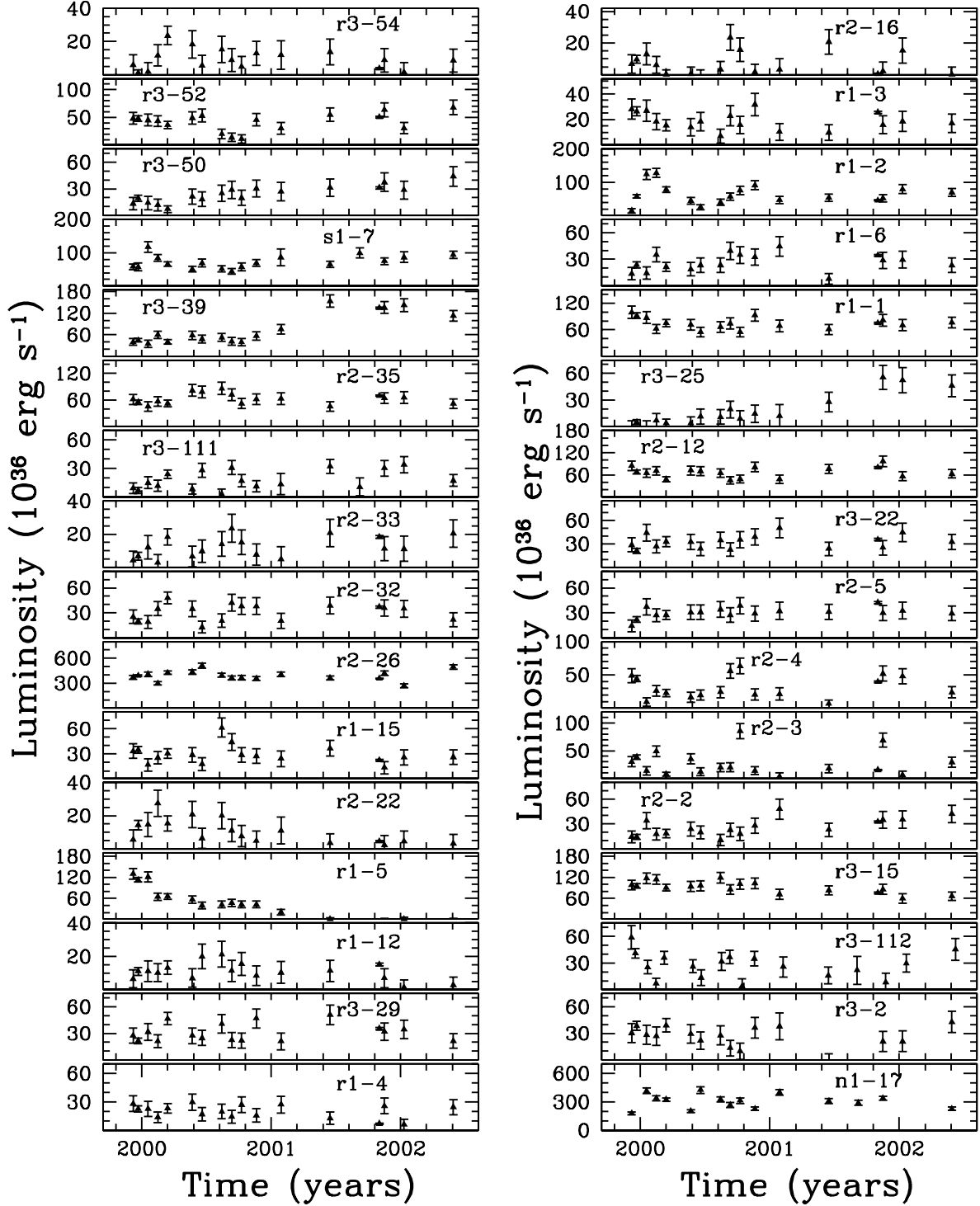,height=8.5in,angle=0}}
\vspace{-0.5cm}
\caption{Long term lightcurves of 32 variable sources objects in M31
are shown.  The other 12 sources: s1-75, s1-79, s1-80, r3-46, r2-28,
r1-34, s1-85, r1-9, r2-67, r3-126, r3-16, and n1-85, are shown in
Figures \ref{nucleus}, \ref{trans}, \ref{stars}, and \ref{143148}.
These sources all have $\chi^2_{\nu}\ \geq\ 1.47$, which provides 90\%
confidence that the sources are intrinsically variable.}
\label{vars}
\end{figure}

In order to better constrain our lightcurves for the objects in the
central region we used the data from \citet{kaaret2002} which covers
18$'$ $\times$ 18$'$ about the center.  This deep observation
provided excellent $S/N$ for the November 2001 data points in our
lightcurves.  Objects near the center of the galaxy therefore show
very small errors for their fluxes during that epoch
(cf. Figure~\ref{vars}).

Farther out in the M31 disk, the density of bright X-ray sources is
small.  This low density of sources allowed larger spatial binning.
This binning was especially useful in the outer parts of the HRC
fields because the effective exposure is lower in these outer portions
and the {\it Chandra} PSF is significantly broader on the outskirts of
the field.  Aside from the lower resolution, the lightcurves were
measured the same way as described for the central region.  The
lightcurves for the variable objects in our sample are shown in
Figures~\ref{vars}, \ref{trans}, \ref{stars}, and \ref{143148}.  These
sources will be discussed further in \S 3.2 and \S 6.

\subsection{ACIS Spectral Analysis of r2-67 and r3-16}

In addition to our HRC analysis of the LF and variability of the X-ray
sources detected in our survey, we analyzed ACIS observations of the
X-ray transients r2-67 and r3-16, the two transients for which we
found counterparts in our HST followup data.  We applied two analysis
techniques to attempt to recover the spectra of r2-67 from the ACIS
observations.  First, we used the pile-up model of \citet{davis2001}
as coded in ISIS 1.0.50 \citep{houck2000}, CIAO 3.0/Sherpa
\citep{freeman2001} and XSPEC V11.2 \citep{arnaud1996}.  Second, we
extracted counts only from the wings of the PSF which are not piled-up
due to their lower counting rates.  Each technique has limitations as
described in the Appendix.  Fitting the spectrum of r3-16 was more
straight-forward, as it was not piled-up.  The spectrum was only fit
using CIAO 3.0/Sherpa.

In all cases we corrected the instrumental response for the
contamination build up on the ACIS detectors, and we limited our
analysis to the 0.3--0.7~keV range where the background is low and the
calibration is secure.  Counts were grouped into bins containing
$\gap$15 counts to allow standard $\chi^2_{\nu}$ statistics, and error
ranges are 68\% as determined from $\chi^2_{\nu}$ contours.  The
details of the fitting procedures are provided in the Appendix.  The
fitting results of r2-67 are discussed in detail in \S 6.3, and those
of r3-16 are discussed in \S 6.4.

\section{X-ray Results}

\subsection{Luminosity Functions}

\begin{figure}
\centerline{\psfig{file=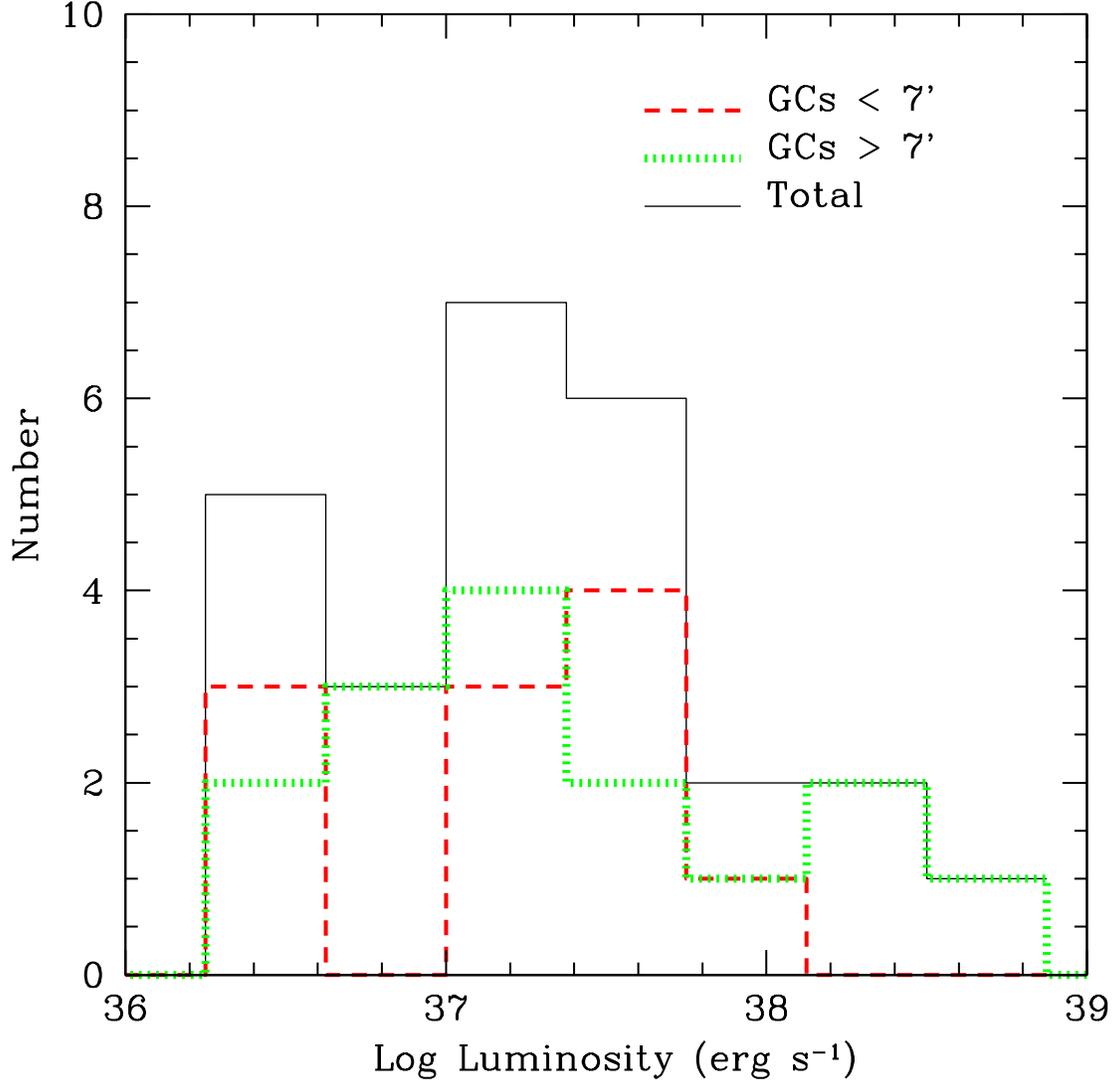,height=6.0in,angle=0}}
\caption{The differential luminosity functions (DLFs) of the GC
sample.  The solid line shows the full GC sample.  A breakdown into
sources near to (dashed line) and far from (dotted line) the center of
M31 is also shown.  A single power-law fit to the the CLF of the GC
population to a minimum luminosity of 1.3$\times$10$^{37}$ \ergs\
yields a slope of 0.84$\pm$0.03.}
\label{gcdlfs}
\end{figure}

In order to look for differences between the disk and bulge source
populations, we measured the LF of the source population within 5
arcmin of the nucleus (the bulge) and outside 7 arcmin of the nucleus
(the disk).  To facilitate comparisons to previous work (i.e.,
\citealp{kong2003,trudolyubov2002,kaaret2002}), we excluded GCs from
our disk sample, and we generated LFs for the bulge both excluding and
including globular clusters within 5 arcmin of the nucleus.  We also
provide the globular cluster luminosity function (GCLF;
cf. Figure~\ref{gcdlfs}).  We always excluded 3 sources likely to be
foreground stars and 2 sources associated with M32 (see
Table~\ref{slist}).  X-ray fluxes were converted to luminosities using
the conversion described in \S 2.2.

\begin{figure}
\centerline{\psfig{file=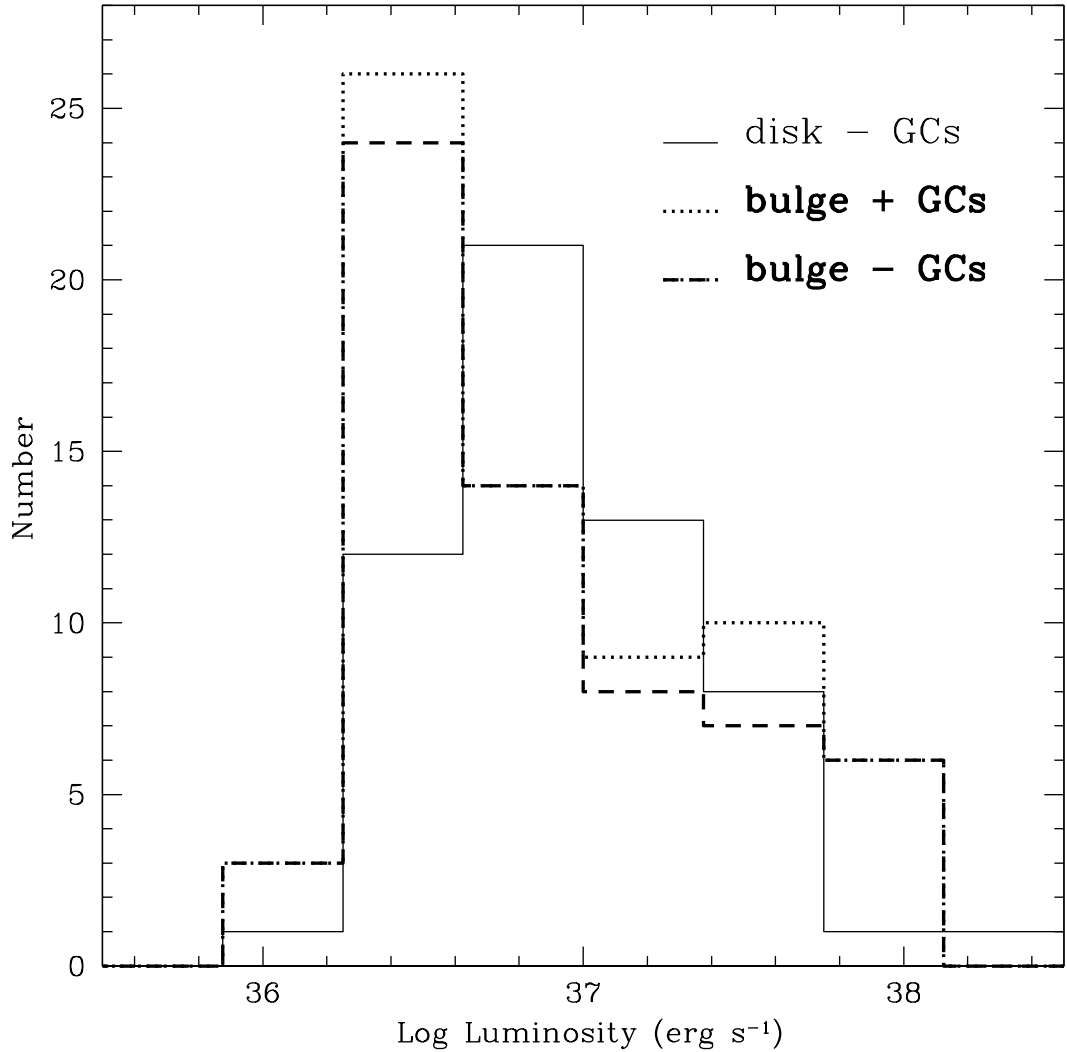,height=6.0in,angle=0}}
\caption{The differential luminosity functions (DLFs) of the disk
(solid histogram) and the bulge (dashed histogram) are shown.  Though
the brightest sources tend to lie in the bulge, the disk sample shows
several sources with luminosities above 10$^{37}$ erg s$^{-1}$.  The
bulge data also reach fainter luminosities.}
\label{dlfs}
\end{figure}

The differential luminosity functions (DLFs) of the disk and bulge
(cf. Fig.~\ref{dlfs}) show some interesting differences.  The LFs of
the bulge with and without GCs are statistically equivalent
(cf. Table~\ref{bplfits}).  The difference in completeness between the
disk and bulge samples is evident.  The bulge sample contains a large
number of sources at $\sim$2.5$\times$10$^{36}$~\ergs , faintward of
which the DLF falls off steeply, revealing the depth of the data.  In
the disk, the DLF falls off in a similar fashion at
$\sim$4$\times$10$^{36}$~\ergs , reflecting the shallower depth in the
disk observations.  This difference in completeness is due to the
variable PSF of {\it Chandra}, which is smaller in the bulge region
therefore allowing fainter sources to be detected.

We created $S/N$ histograms for our source list, revealing a peak in
the number of sources with 6$\sigma$ detections.  We therefore
consider our sample complete for detections of 6$\sigma$ and higher.
A source near the center of the bulge with a luminosity of
2.5$\times$10$^{36}$~\ergs\ will be detected at 6$\sigma$, while a
luminosity of $\sim$3.5 $\times 10^{36}$~\ergs\/ is required for a
6$\sigma$ detection in the disk.  The increase in the {\it Chandra}
PSF with off-axis angle and corresponding drop in sensitivity is
somewhat mitigated by the fact that the field-of-view of the
observations overlaps at the largest off-axis angles, doubling the
exposure time in these regions.  The $6 \sigma$ detection limit is
$3.3 \times 10^{36}$~\ergs\/ in these high exposure regions, but $3.9
\times 10^{36}$~\ergs\/ in nearby non-overlapping areas.  In the
non-overlapping areas farthest off-axis, the $6 \sigma$ detection
limit is $1.3 \times 10^{37}$~\ergs .  For the remainder of the
analysis of the disk LF, we only considered sources with $\rm{L_X} >
4.0 \times 10^{36}$~\ergs, and for the remainder of the analysis of
the bulge LF, we only considered sources with $\rm{L_X} > 2.5 \times
10^{36}$~\ergs .

\begin{landscape}
\begin{figure}
\centerline{\psfig{file=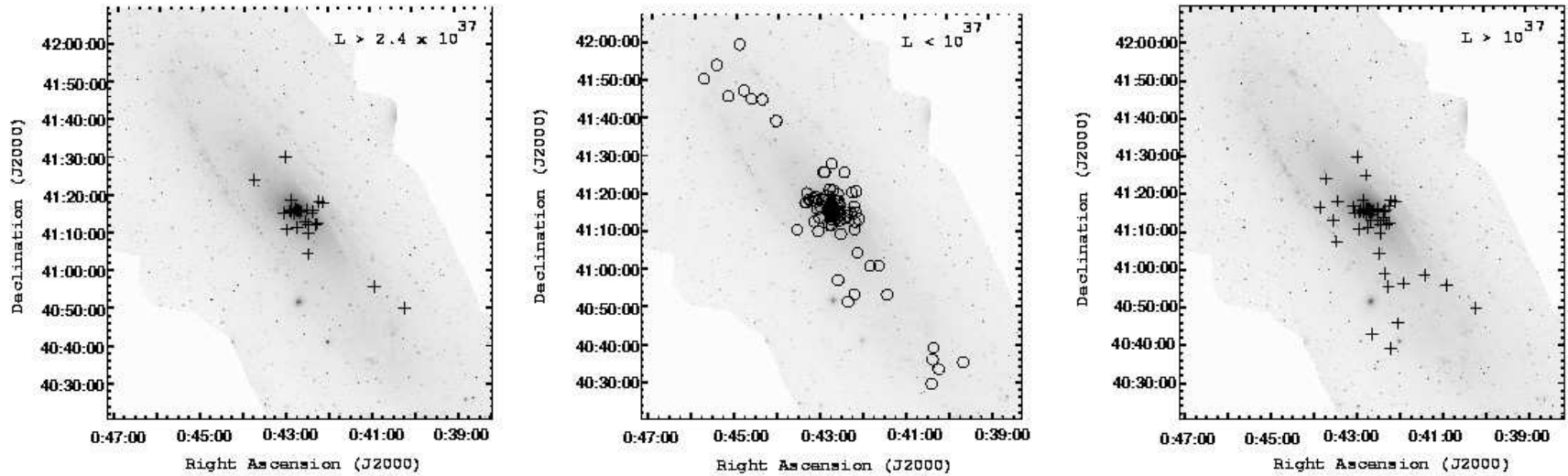,height=2.7in,angle=0}}
\caption{Left panel: the locations of the highest luminosity M31 X-ray
sources ($\gap$2.4$\times$10$^{37}$ \ergs ) are marked with crosses.
Middle panel: the locations of lower luminosity X-ray sources
($\lap$10$^{37}$ \ergs ) are marked with circles.  Right panel: the
locations of M31 X-ray sources with luminosities $\gap$10$^{37}$ \ergs
are marked with crosses, showing there are a number of bright sources
located in the southern disk.  This plot does not include sources in
M32, likely foreground stars, or globular clusters. }
\label{hlll}
\end{figure}
\end{landscape}

Interestingly, the lack of disk sources with luminosities
$\gap$10$^{37}$~\ergs\/ is not as pronounced as seen in the XMM survey
of the northern disk \citep{trudolyubov2002}.  While there is
certainly a decrease in the number of bright sources with distance
from the center of M31, our study hints at an additional, more subtle
effect.  The southern half of the disk contains a large fraction of
the most luminous disk sources.  These sources are shown in
Figure~\ref{hlll}, which shows objects with luminosities
$>10^{37}$\ergs\/ as crosses in the right panel, and objects with
luminosities $<10^{37}$\ergs\/ as circles in the center panel.  The
effect discovered by \citet{trudolyubov2002} is apparent: there are
very few bright sources in the northern disk.  However, the same panel
shows several bright sources in the southern disk.  There is also a
hint of this effect in the data set of \citet{kong2002b} who compared
several regions of the disk.  Their field 2, which lies in the
southern disk, contains most of the bright sources in their sample as
well, but this field was also located closest to the galaxy center.
Of the 9 bright sources south of 41 degrees Declination, tests to a
steady source find $\chi^2_{\nu} >1.47$ only for 3 of them; the rest
have $\chi^2_{\nu} < 1.03$.

\begin{figure}
\centerline{\psfig{file=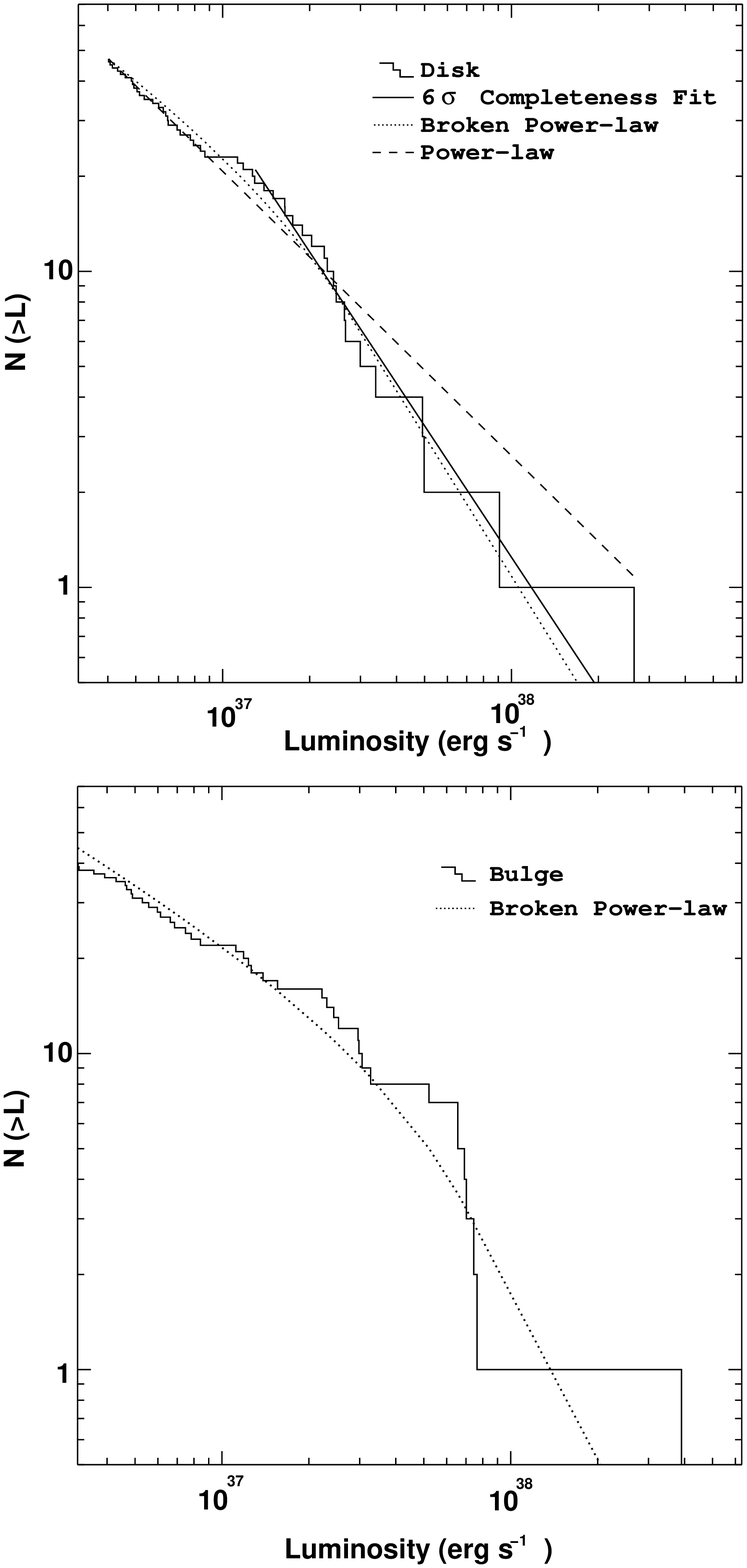,height=6.0in,angle=0}}
\caption{The cumulative luminosity functions (CLFs) of the disk (top
panel) and the bulge (bottom panel).  Both samples were fit with a
broken power-law model (dotted line), where $N(>L) \propto
L^{-\alpha}$.  The broken power law for the disk has $\alpha$ =
0.6$\pm$0.3 below the break and $\alpha$ = 1.5$\pm$0.5 above the break
with the break at $2.6^{+2.5}_{-0.9}\times$10$^{37}$ \ergs.  A single
power-law fit to the disk CLF is also shown (dashed line; $\alpha$ =
0.9$\pm$0.1).  The bulge broken power-law fit has $\alpha$ = 0.5$\pm$0.2
below the break and $\alpha$ = 1.7$\pm$0.7 above the break with the
break at 7.0$^{+2.7}_{-1.3}\times$10$^{37}$ \ergs .}
\label{clfs}
\end{figure}

The cumulative luminosity functions (CLFs) of the disk and bulge are
shown in Figure~\ref{clfs}, and the results of the broken power-law
fits are provided in Table~\ref{bplfits}.  The bulge CLF appears
qualitatively more complex than that of the disk, but it is adequately
fit by a broken power-law with a break at about 7.0$^{+2.7}_{-1.3}$
$\times$ 10$^{37}$\ergs.  This break is higher than the typical values
of $\sim$2.1$\ \times\ 10^{37}$\ergs\ seen in previous surveys
(e.g. \citealp{primini1993,kong2002}).  Our luminosities are 0.1--10
keV while the luminosities of those surveys were for narrower energy
ranges (e.g., 0.3--7 keV; \citealp{kong2002}); our wider energy range
accounts for only $\sim$40\% of the discrepancy, assuming a typical
spectrum ($\alpha=1.7$, $\rm N_H=10^{21}$ cm$^{-2}$).  Our break
luminosity is more easily compared to that measured by
\citet{kaaret2002} which was measured with luminosities of the same
energy range by applying the Maximum Likelihood technique to deeper
(50 ks) HRC data.  Our CLF break measurements agree at the
$\sim$1$\sigma$ level with his measurement of $4.5^{+1.1}_{-2.2}\
\times\ 10^{37}$\ergs.  Our Maximum Likelihood fit broken power-law
has a sharp break, with slope indices of 1.7$\pm$0.7 and 0.5$\pm$0.2
above and below the break, respectively.  These values for the slopes
are also consistent with those measured by \citet{kaaret2002}. The
measured values for these parameters were statistically unaffected by
the presence of GCs in the sample.  The fit is shown on the CLF in the
bottom panel of Figure~\ref{clfs}; this CLF contains no GCs.  Monte
Carlo tests show that 50\% of samples taken from such a broken
power-law distribution provide better fits than our sample.

The disk sample also shows a possible broken power-law consistent with
previous observations \citep{kong2002}.  A Maximum Likelihood fit to
these data gives slopes of 0.6$\pm$0.3 below the break and 1.5$\pm$0.5
above the break with the break at $2.6^{+2.5}_{-0.9}\times$10$^{37}$
\ergs .  This fit is shown in the top panel of Figure~\ref{clfs}.
This distribution is comparable to the inner disk sample of
\citet{kong2002}.  Monte Carlo tests show that 56\% of samples taken
from such a broken power-law distribution provide better fits than our
sample.  The best-fit single power law (cf. Fig.~\ref{clfs} top panel;
Table~\ref{plfits}) has slope 0.9$\pm$0.1.  Monte Carlo tests show
that 71\% of samples taken from such a power-law distribution provide
a better fit than our sample. While the broken power-law fit is
better, both fits are adequate.  Our most conservative CLF for the disk
only includes sources brighter than 1.3$\times$10$^{37}$~\ergs .
Above this luminosity, the entire region of our survey is complete.
This sample is well-fit by a single power-law with index 1.4$\pm$0.2,
also shown in Figure~\ref{clfs}.  Monte Carlo tests show that only 7\%
of samples taken from such a power-law distribution provide a better
fit than our sample.  This index is similar to the value of
1.3$\pm$0.2 found for the northern disk by \citet{trudolyubov2002} to
a faint limit of 10$^{36}$ \ergs .

The higher break luminosity, and slightly flatter CLF below the break
of the bulge is consistent with the brightest sources being in the
bulge.  The steeper CLF in the disk was also seen by \citet{kong2002};
however, their sample did not extend more than 9 arcmin from the
galaxy center.  In contrast, this disk sample contains sources from 7
to 72 arcmin from the nucleus, showing that the steep CLF extends far
out into the disk and the bulge contains most of the bright sources.

Our disk CLF can be compared to that seen in the spiral galaxy NGC
6946 in a survey by \citet{holt2003} with a similar sensitivity limit
($\sim$10$^{37}$ \ergs ).  Their sample is clearly disk dominated, as
the source distribution traces the spiral arms of the galaxy.
\citet{holt2003} found that the CLF of NGC 6946 is a well-behaved
power-law with slope 0.68$\pm$0.03.  The slope in NGC 6946 is
consistent with the recent conclusion by \citet{colbert2003}, from a
sample of X-ray point sources in 32 galaxies, that the CLFs in
late-type spiral galaxies have slopes of 0.5--0.8.  A steeper slope is
seen in the M31 disk in both the large (0.9$\pm$0.1) and the most
conservative (1.4$\pm$0.2) samples.  The slope difference may reflect
the difference in star formation rates of the galaxies, which are
$\sim$4 and $\lap$1 M$_{\odot}$ yr$^{-1}$ in NGC 6946
\citep{sauty1998} and M31 \citep{williams2003b}, respectively.  The
lower current star formation rate in the M31 disk may not replenish
its short-lived, high-luminosity X-ray sources.  Such a process has
been shown to be responsible for steeper CLFs in X-ray population
models \citep{kilgard2002}.

It is interesting to note that the disk CLF for our most conservative
sample is similar to the typical slopes found in elliptical galaxies
($\sim$1.4) in the \citet{colbert2003} sample.  The only one of these
early-type galaxies with a measured star formation rate is NGC 5128
(1.7 M$_{\odot}$ yr$^{-1}$), which has a CLF slope of 1.28.  Because
elliptical galaxies typically have little or no current star
formation, the rate measured for NGC 5128 may be taken as an upper
limit for the other early-type galaxies in the \citet{colbert2003}
sample.  Then the CLF slope and star formation rate of the M31 disk
are typical of what is seen in the early-type galaxies of their
sample.  \citet{colbert2003} were not able to remove GC sources from
their sample or break down the sources into bulge and disk
populations.  Ideally, we would like to compare the M31 disk-only
sample to disk-only samples of more distant galaxies.  Assuming that
the \citet{colbert2003} disk galaxy samples are dominated by disk
sources, the slope of the X-ray CLF of the M31 disk bears a stronger
resemblance to those of typical ellipticals than to those of typical
spirals. 

The GCLFs of our sample are shown in Figure~\ref{gcdlfs}.  The figure
shows the DLF of all GCs in the sample as well as a break-down of the
sample into GCs near the center of M31 and farther out in the M31
disk.  The outer GC sample contains the brightest objects in the
survey.  A power-law fit to the total GCLF down to
1.3$\times$10$^{37}$~\ergs\  yields a slope of 0.84$\pm$0.03
(cf. Table~\ref{plfits}).  Only 0.4\% of 10000 Monte Carlo tests
provided a better fit to this slope than our GC sample.

It should be noted that our LFs for M31 could be affected by the long
baseline and short exposure times of our data set.  For example, a
typical transient is likely to be detected at near its peak luminosity
in one (or a few) exposures, and undetected in the majority of the
remaining exposures. In the extreme case where it is detected at its
peak in a single exposure, and undetected in all other exposures, the
mean luminosity (used in constructing the LF) is 1/17 of peak.  Given
that the duty cycle of transients is likely $\sim$1\%, this extreme
example overestimates the mean luminosity by a factor of $\sim$6.  The
more typical observational mode, consisting of a single long exposure,
may contain only a single (or few) transient(s) at intermediate
luminosity, but our survey detected over a dozen transients.  This
bias will be most severe for sources in the bulge, where the majority
of the transients are located.  To test the effect of the transients
on the CLF of the bulge, we created a bulge LF with the transient
sources removed.  The results of a broken power-law fit to this CLF is
also given in Table~\ref{bplfits}, and is consistent with the fits to
bulge samples both including and excluding GCs.

\subsection{Variability and Transients}

\begin{figure}
\centerline{\psfig{file=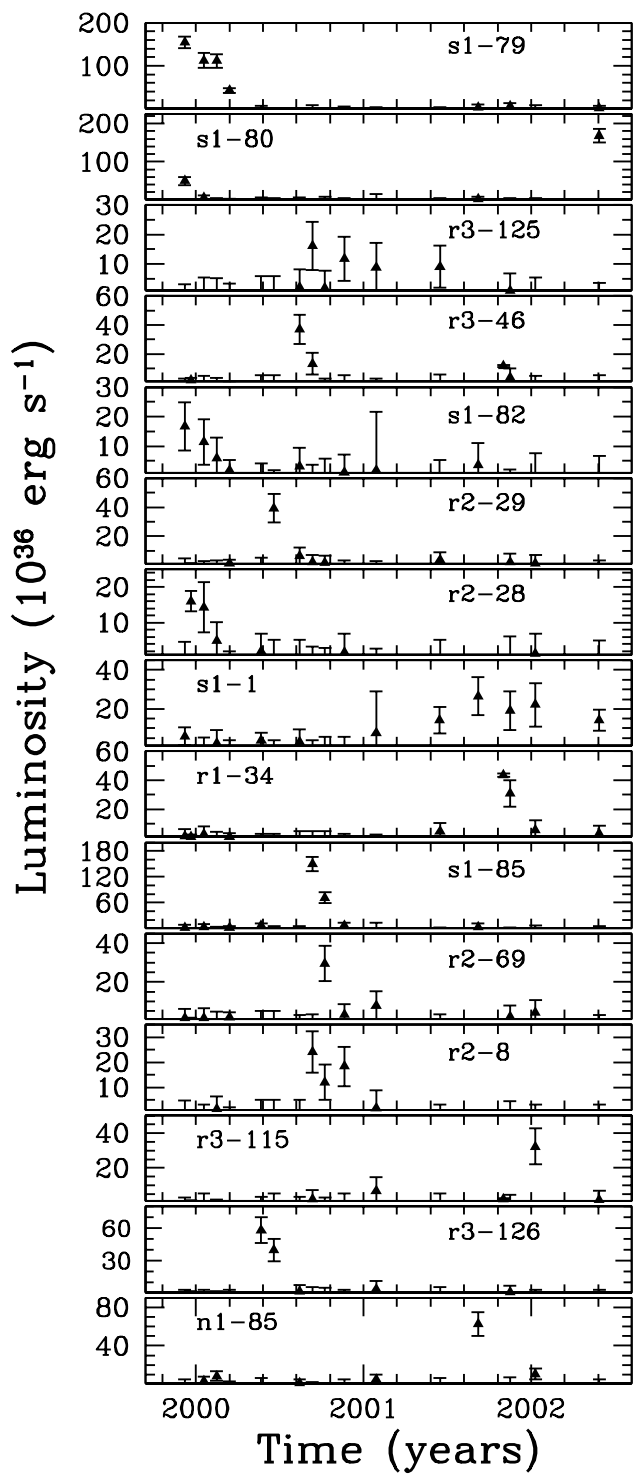,height=8.5in,angle=0}}
\caption{The lightcurves of most of our X-ray transient candidates are
shown.  The lightcurves for the remaining transient candidates, r2-67,
r3-16, and n1-59, are omitted from this figure as they are shown in
Figures \ref{stars} and \ref{143148}.}
\label{trans}
\end{figure}

Any source for which a fit to a constant source at the mean flux
yields $\chi^2_{\nu} > 1.47$ has a 90\% probability of being variable
in our 17 epoch survey. When fitting a model to a data set with 16
degrees of freedom ($\nu = 16$), a value of $\chi^2_{\nu} = 1.47$
leaves a 10\% chance that the model is the correct representation of
the data set (see \citealp{bevington} for details).  There are 44
objects with $\chi^2_{\nu} > 1.47$ in our sample.  The lightcurves for
these objects are shown in Figures~\ref{vars}, \ref{trans},
\ref{stars}, and \ref{143148}.  Among these 44 are 1 (r1-9) quasisoft
source (QSS) and 1 (r2-12) supersoft source (SSS), as diagnosed by
\citet{distefano2003}, and 9 transient candidate objects.  There are
also 8 transient sources with $\chi^2_{\nu} < 1.47$; one of these
(r3-115) is also a SSS.

We defined transient candidates as those objects whose luminosities
reached 1.5$\times$10$^{37}$ \ergs\/ for at least one epoch, but less
than 6 epochs.  These objects were also required to have luminosities
consistent with 0 in at least 2 epochs.  All of the timebins where the
luminosity was consistent with 0 were combined to measure upper limits
of these objects during quiescence.  We also required a 1$\sigma$
upper limit below 2.5$\times$10$^{36}$ \ergs\/ during the combined
quiescent epochs.  The peak luminosities and 1$\sigma$ upper limits of
the quiescent luminosities for the transient candidates are listed in
Table~\ref{tlist}.  Some of the candidates have low $\chi^2_{\nu}$
values due to large errors in the unbinned quiescent epochs.
 
A few of these transient candidates have been detected in previous
surveys.  Object r3-125 was detected by \citet{primini1993} at a
luminosity of $\sim$10$^{37}$\ergs , but this object did not appear in
ACIS observations with a detection limit of $\sim2\times10^{35}$\ergs
, confirming its transient nature.  Objects r3-126 and n1-85 have been
seen in previous surveys and are known to be repeating transients
\citep{osborne2001,white1995}.  Object r3-115 has been recently
identified as a supersoft source (SSS) by \citet{distefano2003}; this
is the only transient source in our catalog corresponding to an SSS in
their catalog.  The position of n1-85 is coincident with a known SNR
candidate.  While the high variability rules out the possibility of
this source being the SNR itself, the source could be associated with
the SNR.  Objects r1-5, r2-16, r2-3, r2-63, and r3-16 have been
classified as transients in previous surveys
\citep{kong2002,distefano2003}; however, they were active too often
during our survey to meet our criteria.  Figure~\ref{active} shows a
histogram of the number of active transients for each epoch of the
survey.  On average there are 2$\pm$1 active transients in M31
during any given epoch.  A new (or recurrent) transient becomes active
about once every 1--2~months. 

\begin{figure}
\centerline{\psfig{file=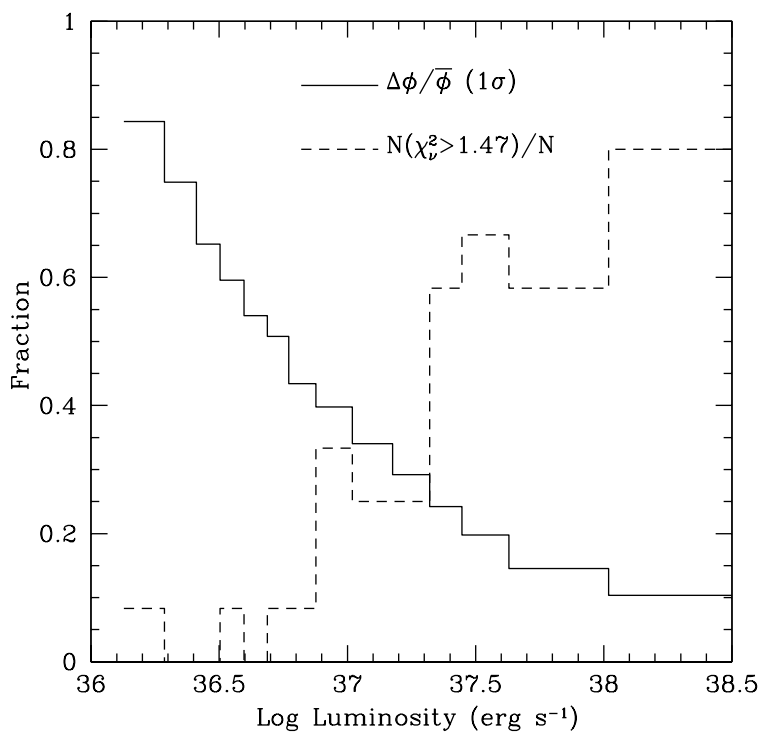,height=4.0in,angle=0}}
\caption{This solid histogram shows the fractional change in X-ray
flux necessary for a 1$\sigma$ deviation in our data set as a function
of source luminosity.  The dashed histogram shows the percentage of
sources with $\chi^2_{\nu}>1.47$ as a function of luminosity.  The
fraction of variables detected rises significantly as the fractional
deviation necessary to detect variability decreases, showing that
many of the faint objects in the sample could have undetected
variability.}
\label{varsen}
\end{figure}

It is interesting to note that half of our transient candidates did
not fit our variability criteria, highlighting the fact that the
fraction of variables we quote truly is a lower limit.  This fact is
made clear in Figure~\ref{varsen}, which shows the fraction of
variables, and the fractional change in flux necessary to show a
1$\sigma$ change, as a function of luminosity in our data set.  Forty
of the 44 variables had mean luminosities greater than
7$\times$10$^{36}$ \ergs , where a flux change of more than 40\% could
be detected.  Half of our sources were below this luminosity.
Therefore, about half of the sources for which our data set is
sensitive to variability at the $\sim$40\% level ($\rm{L_X} >$
7$\times$10$^{36}$ \ergs ) are variable.  It is entirely possible that
half of the sources with mean luminosities $\leq$7$\times$10$^{36}$
\ergs are also variable at the $\sim$40\% level.

The X-ray transient population of M31 is discussed in further detail
in \S 6, including a discussion of the X-ray spectral properties of 2
X-ray transient sources, r2-67 and r3-16, as measured with ACIS.

\section{Optical Observations}

During the course of our survey, wide-field data from the Local Group
Survey (LGS) covering most of the M31 disk became available
\citep{massey2001}.  In addition, new optical data were obtained by
HST through coordinated observations to search for optical
counterparts of five new bright X-ray transient events.  We report on
the detection of 2 counterparts in \S 6.3 and \S 6.4, and will report
on the 3 non-detections in a future paper.  In this section we detail
our analyses of the LGS and HST data and report the results of our
search for potential optical counterparts to all the HRC X-ray
sources.

\subsection{Local Group Survey Data}

New optical data used for this project were generously supplied by the
CTIO/KPNO Local Group Survey (LGS) collaboration
(\citealp{massey2001};\\ http://www.lowell.edu/users/massey/lgsurvey)
which is acquiring 1 arcsec resolution, photometric data with the 8 k
$\times$ 8 k Mosaic cameras on the 4-m telescopes at KPNO and CTIO,
entirely covering ten local group galaxies in $UBVRI$, and narrow-band
H$\alpha$, [S~II] ($\lambda\lambda$6717, 6731) and [O~III]
($\lambda$5007).  The LGS is working on their own, more rigorous,
calibration of these data leading towards a complete $UBVRI$ catalog
of stars.  However, for the purposes of this paper, we have simply
used photometry from the literature to perform a rough calibration.
The analysis used in this paper is described in full detail, including
tests of the photometry routine, in \citet{williams2003}.  In short,
the data consisted of 7 fields from the MOSAIC camera on the KPNO 4-m
telescope.  Observation dates are shown along with the HST observation
dates in Tables \ref{r267obs} and \ref{r316obs}.  These fields cover
most of the active portions of the M31 disk, but do not cover the
bulge.  In order to look for bright, blue stellar counterparts to the
X-ray sources, the Johnson $B$ and $V$ broadband images were analyzed.
The dithered frames in each filter were stacked and reduced using the
DAOPHOT II and ALLSTAR packages \citep{stetson}, and the zero points
for each field were determined using published photometry from
previous surveys of \citet{mochejska2001} and \citet{magnier1992}.

\subsection{HST Data}

Optical data for the X-ray transients r2-67 and r3-16 were obtained
within weeks of the X-ray outbursts through the F336W ($U$-band
equivalent) filter of the WFPC2 camera aboard the {\it Hubble Space
Telescope} (HST).  Eight exposures of 500 seconds each were taken at 3
separate visits for each of the 2 fields observed.  Observation dates
and exposure times are listed in Tables \ref{r267obs} and
\ref{r316obs}.  These 8 images from each visit were analyzed using the
automated photometry package HSTPHOT \citep{dolphin2000}, which is
optimized for the processing and photometric measurement of
undersampled CCD images like those of WFPC2.  The package masks out
all known bad pixels in the field as well as hot pixels flagged by
their deviations from the measured HST PSF.  The images are then
corrected for minor misalignments between exposures and combined.  The
combined images are searched for all source detections with a
$S/N$ greater than 4.  The quality of the PSF fit to each
source is measured to distinguish blends, and the WFPC2 magnitudes are
measured by correcting for the charge transfer efficiency and applying
the photometric calibration of \citet{holtzman1995}.

We applied this analysis method to the 8 exposures from the 3
different epochs for each transient observed.  The combined final
images of each epoch for each object are shown in Figure~\ref{hstims}.
The detections of the UV counterparts of the X-ray transients via
HSTPHOT was therefore objective and provided the F336W ($U$-band
equivalent) magnitudes and errors listed in Tables \ref{r267obs} and
\ref{r316obs}.  The implications of these measurements are discussed
in \S 6.3 and \S 6.4.

\section{Optical Results}

\subsection{Literature Counterparts}

The range in optical magnitudes to be expected for counterparts of
X-ray sources in M31 can be estimated by scaling the optical
magnitudes of the counterparts of Galactic sources.  Galactic
high-mass X-ray binaries (HMXBs) typically contain stars of spectral
type O8--B3 \citep{liu2001}.  These stars have -5$<$M$_V$$<$-2, or
19.8$<$$V$$<$22.8 when scaled to the distance (780 kpc) and reddening
($A_V=0.3$) of M31.  Galactic low-mass X-ray binaries (LMXBs)
typically have -5$<$M$_V$$<$5 \citep{vanparadijs1994}, or
19.8$<$$V$$<$29.8 when scaled to M31.  These estimates suggest that we
may find optical counterparts for HMXBs and bright LMXBs by searching
currently available wide-field surveys, which go to a depth of $V
\approx 23$ (e.g. \citealp{mochejska2001,williams2003b}) outside of
the bulge.  Only 17 stellar counterpart candidates have been found for
our sample.  The small number of high-mass stellar counterparts in the
disk is consistent with the disk XLF, which suggests that the M31 disk
contains very few young bright X-ray sources (cf. \S 3.1).  Only such
young sources are likely to have short-lived, high-mass stellar
companions.  The small number of optical counterpart candidates found
in the bulge is consistent with the effects of crowding in
ground-based surveys of the bulge, which limit the depth to $V \approx
20$ (see upper limits for bulge sources in Table~\ref{lgslist}).
Finding counterparts for most X-ray binaries in the bulge will require
deeper, higher resolution, optical data.

Some HMXBs are transient X-ray sources.  The broad-band optical flux
of these objects is dominated by the high-mass donor star and varies
by $\sim$0.4 mag
(e.g. \citealp{larionov2001,negueruela2001,pigulski2001}).  Even this
low-amplitude variability in the broad-band optical flux may not be
correlated with the variability of the X-ray flux
(e.g. \citealp{negueruela1998,clark1999,larionov2001}).  The optical
magnitudes of these sources are therefore given by the magnitudes of
the high-mass companions, which are typically Be stars with spectral
type O8.5--B2 \citep{negueruela1998}.  These stars have optical
magnitudes of -4.8$<$M$_V$$<$-2.4, or 20.0$<$$V$$<$22.4 when scaled to
M31.

Some LMXBs are transient X-ray sources.  V404~Cyg, one of the most
optically luminous Galactic X-ray transients, had an outburst with
$V$=12.7, $B-V=1.5$, $U-B=0.3$ and a reddening of $A_V=3$
\citep{liu2001} at a distance of 3 kpc
\citep{casares1993,shahbaz1994}.  Scaling to M31, we would expect
$U$=22.4, $B$=22.7, and $V$=22.1.  An example of an optically fainter
Galactic X-ray transient LMXB is the short-period system A0620-00.
Such short-period systems are less luminous than long-period systems
such as V404~Cyg \citep{vanparadijs1994}.  The outburst of A0620-00
had $V$=11.2, $B-V=0.2$, $U-B=-0.8$ and $A_V=1.2$ \citep{liu2001} at a
1.2~kpc distance \citep{gelino2001}.  Scaling to M31, we would expect
$U$=23.4, $B$=24.3, and $V$=24.4.  Assuming these two Galactic
examples bracket the typical optical luminosity range for X-ray
outbursts in LMXBs, the expected range in the visual magnitude of
LMXBs in outburst in M31 is 22.1$<$$V$$<$24.4.  While bright low-mass
transients should be detectable in outburst, they are $>$5 mag
fainter, and therefore undetectable, in quiescence.

In addition, typical GCs in M31 have $V$ magnitudes of 15--19
\citep{barmby2001}, so that GC counterparts are easily identified.
There are also numerous catalogs of emission line sources in M31 that
include potentially X-ray bright objects, such as supernova remnants.
Some of these optical counterparts may also be identified by
cross-correlating the positions in the optical catalogs with the X-ray
source positions.

We searched the literature for previous detections of our sources in
X-rays and at longer wavelengths.  The vast majority of these sources
have no known counterparts at longer wavelengths, as has historically
been the case for X-ray sources in M31.  All but 7 of these sources
have been previously detected in X-rays while only 55 (out of 166) of
them have been detected outside the X-ray band.  Twenty-six of these
are known GCs.  Seventeen are stars from recent ground-based broad
band surveys and the Local Group Survey (cf. \S 5.2).  One is a radio
source classified as a BL Lac candidate.  Three are coincident with SNR
candidates.  One more is a radio source likely to be an SNR.  Six are
classified as planetary nebulae (PNe), but are more likely SNR since
they are X-ray bright, and one is an emission line object of unknown
nature that is optically bright.  Shifting our X-ray positions by
14$''$ and applying an identical search for counterparts yields 0
globular clusters, 3 stars, 3 PNe, and 0 SNRs.  The 14$''$ shift was
larger than the PSF to avoid any real counterparts, but was not so
large as to change the surface density of M31 objects.  The number of
counterparts found for the shifted positions therefore indicates the
expected number of random coincidences between X-ray and optical
sources in our sample.

In some cases, the variability of an object was helpful for
determining the validity of possible counterparts.  For example, r1-15
and r1-2 have coordinates consistent with the coordinates of planetary
nebulae Ford 17 and Ford 13 respectively.  On the other hand, these
sources are unlikely to be planetary nebulae because they are luminous
($\rm L_X > 10^{36}$ \ergs ), and they are unlikely to be SNRs because
they are significantly variable.  These sources are therefore more
mysterious in origin than we may have suspected in the absence of the
long-term variability information.  Four other objects r1-24, r1-23,
r1-26, r3-7 are likely to be SNRs since they have been classified as
PNe and also have constant strong X-ray flux.

\subsection{Local Group Survey Counterparts}

By comparing our X-ray positions to optical data from the Local Group
Survey (LGS) obtained with the MOSAIC camera on the KPNO 4-m at 1$''$
resolution \citep{massey2001,williams2003}, we were able to find
several new optical counterpart candidates.  These candidates are
listed in Table~\ref{lgslist}.  The regions currently surveyed include
$\sim$75\% of the disk, but very little of the bulge, where most of
the bright X-ray sources lie.  Crowding in ground-based images of the
M31 bulge severely limits searches for counterparts, so that
comparisons to future LGS data from the bulge is unlikely to yield
many new counterparts.

Four of these new candidates have appropriate colors and magnitudes to
be foreground stars (s1-74,s1-45,n1-82,n1-59).  Typically, Galactic
foreground stars in the direction of M31 have colors of ${\rm B-V} >
0.4$ with the highest number having colors of ${\rm B-V} \sim 1.6$
\citep{hodgeleemateo1988}.  Many M31 stars have these colors as well,
so that the only way to be sure a star redder than ${\rm B-V} \sim
0.4$ is in M31 is with spectroscopy.  Because such spectra are not
available to us, we assume the stars most likely to be foreground are
bright red stars (${\rm B-V} \gap 0.4$, V$\lap$18).  Of these
foreground candidates, objects s1-74 and n1-82 were too faint to
obtain a reliable measurement of variability, and s1-45 is unlikely to
be variable, with a $\chi^2_{\nu}$ value of 1.25.  Object n1-59 is
especially interesting, as it is a transient candidate.

\begin{figure}
\centerline{\psfig{file=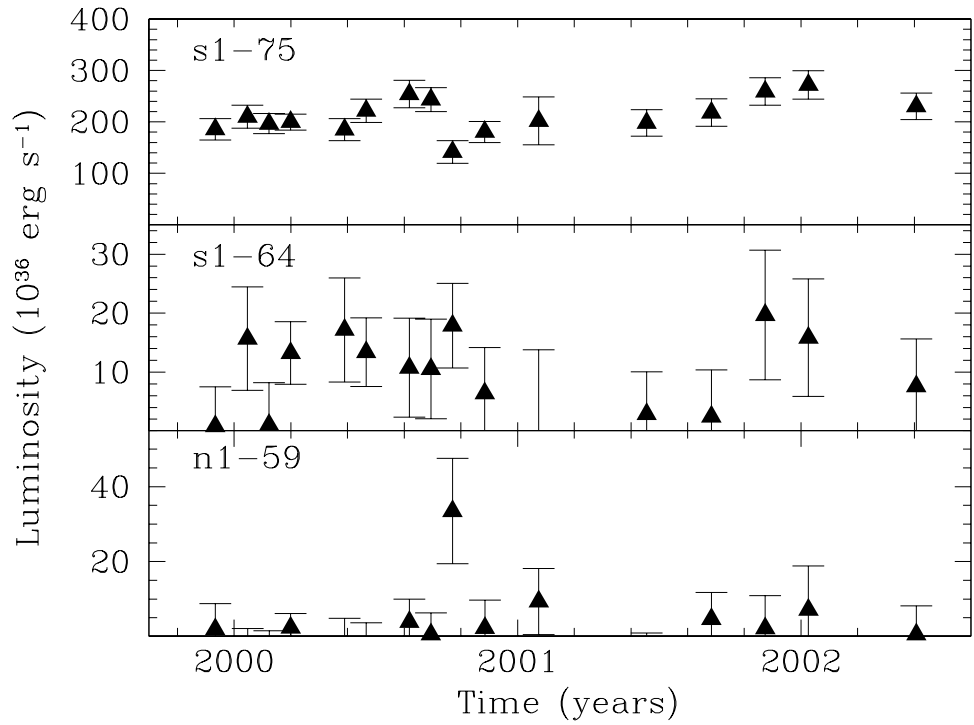,height=4.0in,angle=270}}
\caption{Long term lightcurves of 3 of the brightest X-ray sources
with counterparts in the LGS data.  Object s1-75 is a known BL Lac
candidate, s1-64 has a color and magnitude consistent with being an
M31 member, and n1-59 is an X-ray transient candidate.}
\label{stars}
\end{figure}

On the other hand, there were also 11 counterparts with colors and
magnitudes typical of M31 upper main sequence or giant stars (s1-75,
s1-78, s1-64, r3-28, r2-8, r2-67, r3-115, r3-16, r3-13, r3-7, n1-81).
Object s1-75 is associated with a BL Lac candidate.  Objects r2-8, and
r3-115 have colors appropriate for M31 red giants.  The X-ray
transients r2-67 and r3-16 are discussed in detail in \S 6.3 and \S
6.4.  Objects s1-78, s1-64, r3-28, r3-13, r3-7 and n1-81 have the
colors and magnitudes of evolved and/or reddened M31 upper main
sequence stars.  The counterpart candidates most likely to be in M31
are those with B-V$\lap$0.4 (s1-64, r3-28, r3-13 and r3-7).  Although
these candidates are all slightly $>$1$''$ from their possible X-ray
counterparts, their X-ray positions have errors (given by {\it
wavdetect}) of 0.6$''$, 0.5$''$, 0.5$''$, and 0.4$''$ respectively,
and the LGS coordinates are accurate to about 1$''$, allowing the
possibility that these are the true optical counterparts The majority
of these possible stellar counterparts have too few X-ray photons per
1~ks exposure to generate meaningful lightcurves, but s1-75, s1-64,
and n1-59 are brighter.  The X-ray lightcurves for these 3 are shown
in Figure~\ref{stars}. Tests to a constant source find $\chi^2_{\nu} =
1.97, 0.69$, and 0.68 for s1-75, s1-64, and n1-59
respectively. Variability is therefore detected only in s1-75, the
known BL Lac candidate.

There were also several X-ray sources in regions observed by the LGS
that did not correspond to any point sources in the LGS optical
images.  For these objects, we obtained lower limits for their $V$
band magnitudes by measuring the completeness of the area of interest
as a function of $V$ magnitude.  The lower limit was chosen as the V
magnitude where the mean completeness of the LGS data fell below 50\%.
These limits are provided in Table~\ref{lgslist}.

\section{X-ray Transients}

\subsection{The X-ray Transient Population}

X-ray transient sources located in the bulge (within 7$'$ of the
nucleus) are labeled with a (B) in Table~\ref{tlist}.  These are
likely to be LMXB transients because most stellar population and
interstellar medium studies of the M31 bulge suggest that it is
dominated by old stars and contains very few, if any, young stars
(e.g. \citealp{stephens2003,davidge2001,brown1998,vandenbergh1991m31}).
Indeed, the central 44$''$ contains no stars earlier than B5
\citep{king1992}, and only an upper-limit to the star formation rate
of the bulge has been measured from the far-UV luminosity
($\leq7.4\times10^{-5}\ M_{\odot}\ yr^{-1}$; \citealp{deharveng1982}).
In addition, with deep HRC imaging taken during the outburst of r2-67,
\citet{kaaret2002} found no evidence for the existence of X-ray
pulsars in the M31 bulge, suggesting it lacks HMXBs.

While the possibility of a high mass secondary in the bulge is low, it
is not zero.  The bulge is extremely crowded, and the highest
resolution surveys have not covered the entire region or analyzed
spectra of all bright blue stars.  In addition, the discovery of
molecular clouds \citep{melchoir2000} and supernova remnants
\citep{kong2003,sjouwerman2001} in the M31 bulge allow the possibility
for some recent star formation.  The LGS optical counterpart
candidates for r2-8, r2-67 and r3-115 have absolute $V$ magnitudes
(neglecting absorption) of M$_V \approx -3.5, -2.5$ and $-2.6$,
respectively, consistent with high mass (e.g., Be binary)
counterparts.  The colors measured for these objects from the
ground-based data are not completely reliable, as the $B$ and $V$
observations were separated by nearly 1 year and Be stars are
typically variable \citep{pigulski2001}.  However, we note that apart
from times of major outbursts, the scale of variability is small so
that these colors should be reasonable estimates.  In addition, the
ground-based data are very crowded in these regions which often causes
the optical luminosity to be over-estimated.

Within the disk the star formation rate is higher and there are
numerous high mass stars, so transients within the disk may have high
mass secondaries.  HMXB transients often contain Be secondaries and
pulsing neutron star primaries \citep{tanaka1996}, and are therefore
called Be transients.  Only one optical counterpart was found for a
disk transient: n1-59 in the northern disk.  This source has a very
bright optical counterpart candidate, $V=17.15\pm0.10$.  If this
object is indeed the optical counterpart and the source is in M31,
then M$_V \approx -7.3$ (or even brighter if we include the likely
$\sim$1~mag of reddening), which makes it at least as bright as an
early O star.  Be stars typically have spectral types between O8.5--B2
\citep{negueruela1998} and optical magnitudes of $-4.8 \lap$ M$_V \lap
-2.4$ \citep{cox2000}.  The counterpart candidate for n1-59 is
therefore likely a foreground object.

\subsection{Ratio of Black Hole to Neutron Star X-ray Binaries}

The relative numbers of X-ray binaries containing black hole
vs. neutron star primaries within the M31 bulge may be estimated from
the relative numbers of transients vs. persistent sources.  We follow
the same argument that has been used to estimate this ratio within the
Galaxy \citep{verbunt1995,portegies1997}.

The argument goes as follows.  Within the bulge, the binary population
is dominated by low-mass ($\lap$1 $M_{\odot}$) systems because the
secondaries are old.  Within the Galaxy, persistent LMXBs are observed
to often contain neutron star primaries.  On the other hand, a large
fraction of the transient low-mass systems are observed to contain
black hole (BH) primaries \citep{charles1998}.  The duty cycle of
these transient black hole X-ray novae (BHXNe) is believed to be
$\sim$1\% \citep{tanaka1995}.  Thus the ratio of persistent to
transient sources, divided by the duty cycle of the transients, yields
an estimate of the ratio of number of neutron star containing LMXBs to
the number of black hole containing LMXBs.  Within the Galaxy, it
appears that there are approximately equal numbers of neutron star and
black hole containing LMXBs.  This result is surprising, as
evolutionary calculations predict that neutron star containing systems
should be $\sim$100 times more common than those containing black
holes \citep{portegies1997}.

\begin{figure}
\centerline{\psfig{file=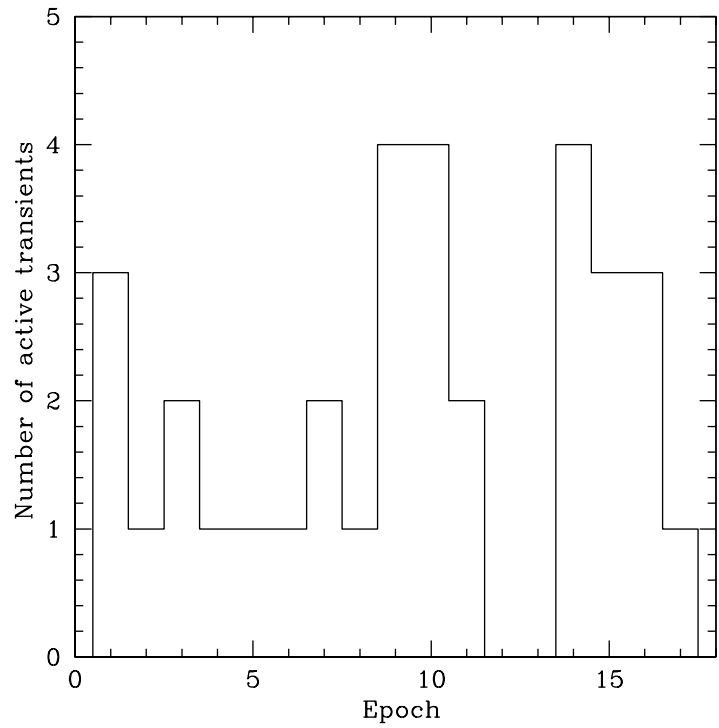,height=4.0in,angle=0}}
\caption{This histogram shows the number of active X-ray transient
sources in M31 during each epoch of the survey.}
\label{active}
\end{figure}

Within the bulge of M31, there are $\sim$100 persistent LMXBs.  In any
individual snapshot, there is on average $\lap$1 new bright transient
within the bulge (e.g., Figure~\ref{active}).  The majority of these
are detected in more than one snapshot, and therefore have decay times
consistent with BHXNe \citep{chen1997}.  If the duty cycles of the
BHXNe in M31 are the same as in the Galaxy (and there is no reason to
suspect otherwise), then the problem observed in the Galaxy is also
seen in the M31 bulge: the number of black hole containing LMXBs is
approximately equal to the number of neutron star containing LMXBs.

\subsection{X-ray Transient r2-67}

\subsubsection{ACIS Observations}

Source r2-67 was observed several times with the ACIS detectors.
Herein, we limit our discussion of the spectra to OBSID 1585 and OBSID
2897.  Spectral analysis of these observations was performed to
estimate the X-ray luminosity of this transient, as well as its
optical extinction.  These values were critical to our determination
of its orbital period.  The other ACIS detections will be discussed in
detail in future publications.  Given the long decay time of this
transient, the X-ray luminosity from OBSID 1585 is close enough in
time to the measurement of the optical luminosity from HST that the
luminosities can be used in the correlation of
\citet{vanparadijs1994}.  Analysis of this observation is complicated
because the source was bright and piled-up (cf. Appendix).  During
OBSIB 2897 the transient had faded and pileup was negligible. When the
transient is faint, spectral fitting is more straightforward, but
given the possibility that the spectrum may be time variable, it is
desirable to measure it contemporaneously with the HST $U$-band
measurements.

\subsubsection{OBSID 1585}

The counting rate during 2001 Nov. 19 (OBSID 1585), at 0.18 \cs , was
high enough to give a pileup fraction of $\sim$30\% (as fit by Sherpa
and ISIS).  In order to apply the pileup model, we follow the
prescription as described by \citet{davis2001,davis2003}.  As this
method is relatively new, we include a detailed discussion of our use
of it in the Appendix.  ISIS quickly converges to a pileup model with
a grade migration parameter $\alpha_G = 0.99$, with a fixed PSF
fraction of 0.95.

Using these pileup parameters, we fit an absorbed disk blackbody model
to the spectrum.  The best fit absorption value was $\rm N_H = 2\pm1
\times 10^{21}$ cm$^{-2}$, corresponding to ${A_V} = 1.1\pm0.6$
\citep{predehl1995} and ${A_U} = 1.7\pm0.8$ for a standard extinction
law with R$ = 3.1$ \citep{cox2000}.  The best fit disk parameters were
$T_{inn} = 0.35\pm 0.05 $~keV and $N = 11^{+20}_{-8} (R_{inn}/km)^2
(10/d)^2 cos (\theta)$, where $T_{inn}$ is the temperature of the
inner edge of the accretion disk, $N$ is the normalization parameter,
$d$ is the distance to the source, and $\theta$ is the inclination
angle of the disk.  Assuming a distance of 780 kpc, $N$ provides a
lower limit to the radius of the inner edge of the accretion disk,
$R_{inn} \gap 140$ km.  These parameters provide a fit with
$\chi^2_{\nu}$ = 1.2 (probability=0.2).  The observed flux is 1.3
$\times$ 10$^{-12}$ \ergcm2s (0.3-7.0 keV), and the modeled emitted
luminosity is $1.9 \times 10^{38}$ \ergs (0.3-7.0 keV).  The $R_{inn}
\gap 140$ km corresponds to $R_{inn} \gap 10 r_g$ for a 10 $M_{\odot}$
black hole, and $R_{inn} \sim 20 r_g$ for the best fit values assuming
cos($\theta$) = 1.  These are reasonable values for an accretion disk
around a $\sim$10 solar mass black hole during this bright stage of
the outburst.

\vspace{-0.5cm}
\subsubsection{OBSID 2897}

By 2002 Jan.~8, the flux had decayed sufficiently that pileup was no
longer a problem.  We extracted 177 counts from a $2''$ radius, and
found that a simple power-law with ISM absorption models fit the data
well ($\chi^2_{\nu} = 1.01$).  The observed 0.3--7.0~keV flux was $2.9
\times 10^{-13}$ \ergcm2s~, and the modeled emitted 0.3--7.0~keV
luminosity at 780~kpc was $2.9 \times 10^{37}$ \ergs .  The best fit
value of ${\rm N_H} = 0.7\pm1.3 \times 10^{21}$ cm$^2$ is consistent
with that due to the Galaxy alone, and corresponds to ${A_V} =
0.4\pm0.7$ \citep{predehl1995} and ${A_U} = 0.6\pm1.1$ for a standard
extinction law with R$ = 3.1$ \citep{cox2000}.  This measurement is
consistent with the measurement from OBSID 1585 of ${\rm N_H} = 2\pm1
\times 10^{21}$ cm$^2$.  When the two measurements are combined, the
best estimate of the absorption to r2-67 is ${\rm N_H} = 1.5\pm0.8
\times 10^{21}$ cm$^2$, which corresponds to ${A_V} = 0.8\pm0.4$.

We note that the best fit slope is substantially harder than that
found when the source was bright (see Appendix for a power-law fit to
the bright state data).  The harder slope is consistent with the
source having entered the `low-hard' or `intermediate' state from the
`high-soft' state in OBSID~1585 \citep{esin1997}.  As these
transitions are believed to occur at luminosities between 10\% and a
few \% of Eddington, this transition is consistent with the source
having a mass of $\sim$10~\msol .

\subsubsection{Counterparts for r2-67}

Two HST observations taken $\sim$1 month apart, both when the X-ray
source was bright, reveal a $U$-band source with F336W = 22.3$\pm$0.1 mag.
A third observation several months later, when the X-ray source was
faint, did not detect the counterpart and set a limit of F336W $>$
22.8 mag.  The disappearance of the optical counterpart in concert with
the X-ray decline confirms this identification of the optical
counterpart of the X-ray transient.  The observations are shown in
Figures~\ref{hstims} and \ref{143148} and summarized in
Table~\ref{r267obs}.

Fortuitously, most of the HST images contained an X-ray bright GC,
which we used to register with our {\it Chandra}-HRC mosaic to the HST
images.  X-ray position errors were estimated by dividing the
full-width at half maximum (FWHM) of the X-ray source, as measured
with the IRAF\footnote{IRAF is distributed by the National Optical
Astronomy Observatory, which is operated by the Association of
Universities for Research in Astronomy, Inc., under cooperative
agreement with the National Science Foundation.} task {\it imexamine},
by the square-root of the number of counts in the X-ray detection.
The final position errors were dominated by the X-ray position errors;
the optical position errors were always $<$0.1$''$.  The first 2 HST
observations of r2-67 were aligned with our X-ray image using the GC
Bol 148.  The final HST observation of r2-67 was aligned using GC Bol
144.  Our {\it Chandra} positions of the GC sources had errors of
0.15$''$ and 0.08$''$, respectively, and the X-ray position of the
r2-67 had an error of 0.2$''$.  Therefore, in Figure~\ref{hstims}, the
0.8$''$ error circles marked on the first two HST images of r2-67, and
the 0.6$''$ error circle marked on the third HST image of r2-67, show
the 3$\sigma$ position errors of our data.

Interestingly, a candidate was also found in the LGS data for r2-67.
This candidate has $B$=22.3$\pm$0.3 and $V$=21.9$\pm$0.1.  The
counterpart for r2-67 seen from the ground was particularly unexpected
because the $B$-band images were taken October 6, 2000, when the X-ray
source was quiescent, and the $V$-band images were taken September 18,
2001, a few months before we detected the transient but during a 3
month gap in our X-ray monitoring.  The LGS object may be the
counterpart or a chance superposition of a different star along the
line of sight, while the $U$-band transient seen by HST is clearly the
optical glow of the X-ray nova.  The magnitude of the LGS star from
the ground is likely a lower limit as crowding often causes the
brightness to be overestimated.

The apparently persistent nature of the LGS candidate (i.e. it was
detected when we believe the X-ray source was dim) and the clear
transient nature of the HST counterpart calls into question the
validity of the LGS candidate.  However, the LGS candidate is
well-detected in the $V$-band, with $S/N$ of 9, but the detection is
less robust in the $B$-band, with $S/N$ of 4.  The X-ray source is not
a pulsar \citep{kaaret2002} and is therefore unlikely to have a
high-mass (Be) companion which would be persistent optically.  While
the LGS and on-state HST magnitudes are approximately equal to those
expected for a slightly evolved $\sim$B3 star in M31 with $E_{B-V}
\approx 0.5$, such an object would have been detected in all 3 HST
observations.  The LGS photometry allows the possibility that this
optical detection is an interloping foreground main-sequence star with
F336W $\gap$22.8 mag and $B-V \approx 0.7$.  This possibility offers
the simplest explanation for the candidate's non-detection in the
third HST image.  The detection of this counterpart candidate in the
LGS data underscores the confusion that crowding causes in
ground-based images of the M31 bulge.

\vspace{-0.5cm}
\subsubsection{${\rm L_X/L_{\rm opt}}$ Determination of the Orbital Period}
 
It is possible to estimate the orbital period for the transient using
the period dependence within the correlation between X-ray and optical
luminosity found by \citet{vanparadijs1994}.  In order to do so,
estimates of the X-ray luminosity and the $V$-band absolute magnitude
are needed.  While the former were directly available from the HRC and
ACIS observations, the latter were estimated from the HST (F336W)
measurements and corrected for the interstellar absorption to the
transient, as estimated from the ACIS data (cf. \S 6.1) and converted
to $A_V$ using the correlation between between ${\rm N_H}$ and $A_V$
(i.e., \citealp{predehl1995}).

Plugging the contemporaneous optical and X-ray luminosities into the
\citet{vanparadijs1994} relation we can estimate the orbital period of
r2-67.  The $A_U$ indicated by combining our X-ray measurements of the
absorption toward r2-67 (cf. \S 6.3.3) is $A_U = 1.3\pm0.7$.  This
$A_U$ implies an absolute M$_U = -3.4$.  We assume an intrinsic $U-V =
-1.0\pm0.4$, typical for observations for LMXBs in the Galaxy
\citep{liu2001}.  These assumptions supply an estimate for M$_V =
-2.4\pm0.8$.  Along with the observed ${\rm L_X} = 1.9 \times 10^{38}$
\ergs, this M$_V$ implies an orbital period ${\rm P_{orb}} =
23^{+54}_{-16}$~days.  This is not unreasonable compared to BHXNe
within the Galaxy.  For example V404~Cyg has ${\rm P_{orb}} =
6.47$~days \citep{orosz2002}, and GRS1915+105 has ${\rm P_{orb}}
=34$~days \citep{greiner2001}.

\vspace{-0.5cm}
\subsubsection{Disk Decay Time Determination of the Orbital Period}

A second way to estimate the orbital period is from the decay time of
the outburst.  \citet{king1998} developed a model for the outburst of
BHXNe disks, assuming that irradiation determines the disk temperature
profile during the outburst and that the outburst cannot end until the
irradiation allows the outer edge of the disk to cool below the
hydrogen recombination temperature. This model predicts longer decay
times for larger disks and approximately linear decay curves for
systems with orbital periods longer than $\sim$1 day.  Shorter orbital
period systems are predicted to produce exponential decay curves.
While the data herein do not constrain the shape of the lightcurve
(cf. Fig.~\ref{143148}), they do provide a decay time estimate of
$\sim$0.2~yr.

Equation 23 of \citet{king1998} describes the time variable mass
transfer rate for such a large disk in outburst. Assuming ${\rm L_X} =
\dot{M}_c \eta c^2$, where $\dot{M}_c$ is the central accretion rate
and $\eta $ is the accretion efficiency ($\sim$0.1) allows us to
re-write this as
$$
{\rm L_X} = \eta c^2(3\nu/B_1)^{1/2}M_h^{1/2} - \eta c^2 (3\nu/B_1)t\ \ \ \ \ \ \ \ \ \ \ \ \ \ (1)\\
$$ 
where $\nu$ is the disk-averaged kinetic viscosity, the constant $B_1
= 4\times 10^5$ (cgs units), $M_h$ is the mass of the hot zone and $t$
is the time in seconds after the start of the outburst decay.

Applying our observation that ${\rm L_X} = 1.9 \times 10^{38}$ when $t=0$
to equation (1) allows us to write
$$
{\rm L_X} = 1.9\times 10^{38} = \eta c^2 (3\nu/B_1)^{1/2}M_h^{1/2}.\ \ \ \ \ \ \ \ \ \ \ \ \ \ \ \ (2)
$$
Using the value in equation (2) as the first term in equation (1),
applying our observation of the decay time of $t=0.2$~years, and
assuming that the luminosity is $\sim$0 at this point (justified since
it is $ \ll 1.9\times 10^{38}$ at $t = 0.2$ years) lets us write
$$
{\rm L_X} \simeq  0 = 1.9\times 10^{38} - \eta c^2 (3\nu/B_1) (6.3 \times
10^6 s).\ \ \ \ \ \ \ \ (3)  
$$

Equation (3) lets us determine the viscosity, $\nu \simeq 4.5 \times
10^{16}$ cm$^2$ s, a value characteristic of the large disk implied by
the $23^{+54}_{-16}$ day period.  The \citet{shakura1973}
alpha-prescription states that
$$
\nu = \alpha c_s H = \alpha c_s^2(R^3/GM)^{1/2}\ \ \ \ \ \ \ \ \ \ \ \ \ \ \ \ \ \ \ \ \ \ \ \ \ \ \ \ \ \ \ (4)
$$
where $\alpha$ is the angular momentum transport efficiency and $H$ is
the disk thickness.  In the thin disk approximation, $H =
c_s(R^3/GM)^{1/2}$ (see \citealp{frank1992}, p. 74).  Assuming $c_s^2 = kT/m_p$, the outer disk radius, where the disk cools below the ionization temperature, has the value of
$$
R_{disk} \simeq 2 \times 10^{12}{M_{10}^{1/3}\over \alpha^{2/3}T_4^{2/3}}\ \rm cm\ \ \ \ \ \ \ \ \ \ \ \ \ \ (5)
$$
where $M_{10}$ is the BH mass in 10 $M_{\odot}$ units, and $T_4$ is the local
disk temperature in units of $10^4$ K. This is consistent with the binary
separation,
$$
a = 5^{+6}_{-3} \times 10^{12} M_{10}^{1/3}\ \rm cm,
$$ 
calculated from the $23^{+54}_{-16}$~day period with Kepler's law.
The similarity of these values suggests that, in addition to the ${\rm
L_X/L_{opt}}$ ratio, the outburst decay time is also consistent with that
of a BHXN with a  $23^{+54}_{-16}$~day orbital period.

\subsection{X-ray Transient r3-16}

\begin{figure}
\centerline{\psfig{file=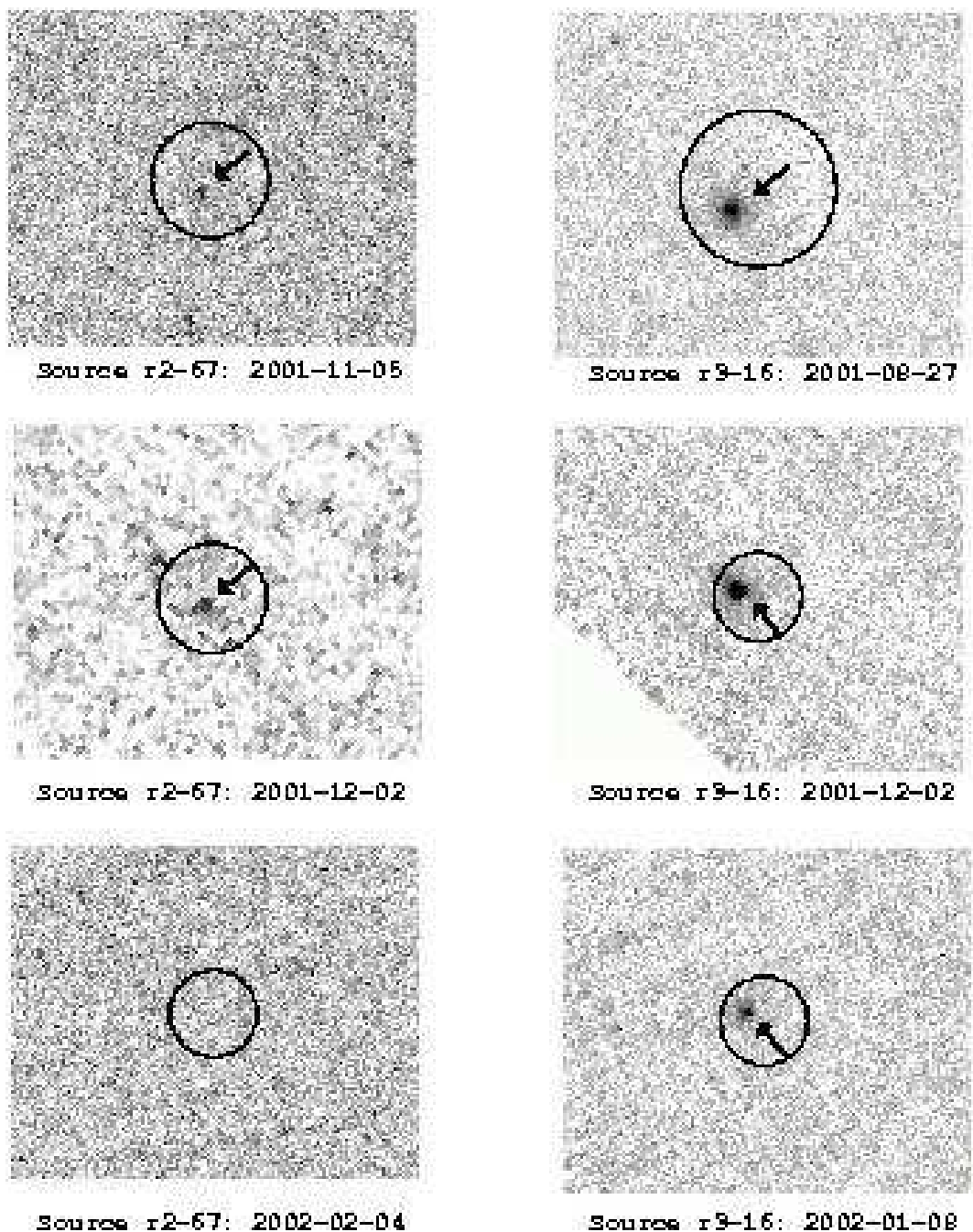,height=7in,angle=0}}
\caption{HST/WFPC2 F336W images of the X-ray transients r2-67 (CXOM31
J004305.6+411703) and r3-16 (CXOM31 J004309.7+411901) are shown.
Overplotted on each image is the 3$\sigma$ error circle for the
position of the X-ray transient detected (position errors are
discussed in \S 6.3.4 and \S 6.4).  These circles have radii of
0.8$''$ in the first two r2-67 images, 0.6$''$ in the third r2-67
image, 2.3$''$ in the first r3-16 image, and 1.3$''$ in the last two
r3-16 images.  Arrows mark the object we claim to be the counterpart
for r2-67, which was observed by HST on three occasions after its
detection October 31, 2001.  It was detected in the $U$ band for two
months, then faded.  Object r3-16 is seen in all optical bands from
the ground, and is also seen in all three HST images.}
\label{hstims}
\end{figure}

\begin{figure}
\centerline{\psfig{file=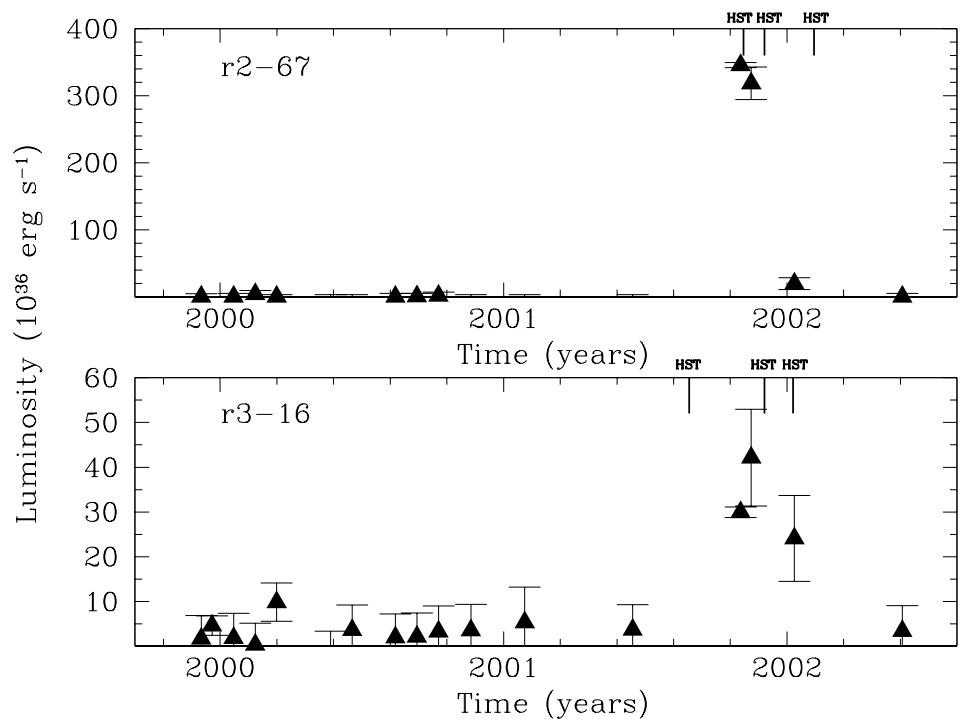,height=4.0in,angle=270}}
\caption{Long term lightcurves of the X-ray transients r2-67 (CXOM31
J004305.6+411703) and r3-16 (CXOM31 J004309.7+411901) are shown.  The
dates of contemporaneous HST observations are labeled with vertical
marks from the top axis.  Object r2-67 was observed by HST on three
occasions after its detection October 31, 2001.  It was detected in
the $U$ band for two months, then faded.  Object r3-16 is seen in all
optical bands from the ground, and is also seen in HST images.  The
first 2 HST images of r3-16 were serendipitous, as the object was
located in the same field as a different transient event.}
\label{143148}
\end{figure}

Another highly variable source, r3-16, whose lightcurve is also shown
in Figure~\ref{143148}, has a bright UV counterpart of unknown nature.
HST images of the optical counterpart during 3 epochs are provided in
Figure~\ref{hstims}.  The second and third HST observations of source
r3-16 were aligned with our X-ray mosaic using the GC Bol 148.  The
error on the X-ray position of r3-16 was 0.4$''$, and that of Bol 148
was 0.15$''$.  The 3$\sigma$ error circle for this source was
therefore 1.3$''$, and it is shown in Figure~\ref{hstims}.  The first
HST observation of r3-16 did not contain any X-ray bright GC sources.
For this observation, we had to rely on the original coordinate system
assigned by the HST pipeline for our alignment.  We added 1$''$ to the
radius of our error circle in this case to account for uncertainty in
the registration between the {\it Chandra} and HST coordinates.
Therefore, the error circle in the first image has a 2.3$''$ radius.
The F336W magnitudes from HST observations and $BV$ magnitudes from
ground-based observations are provided in Table~\ref{r316obs}. The
first 2 HST observations were taken near the peak of the X-ray
outburst, and the third during the decline.  The HST F336W magnitudes
are $\sim$21 and appear variable, while the LGS data yield
$B$=19.1$\pm$0.1 and $V$=18.8$\pm$0.1.  

Discerning the nature of object r3-16 is difficult without any
high-quality optical spectral information.  The object was classified
as an emission line object of unknown nature by \citet{wirth1985}.
However, the X-ray spectrum and optical size of the object provide
some new hints about its nature.  We extracted the spectrum from a
January 8, 2002 ACIS observation (OBSID 2897) and found that it is
well fit with an absorbed power-law with ${\rm N_H}$ = 1.8$\pm$0.8
$\times$ 10$^{21}$ cm$^{-2}$ and slope $\alpha=1.9\pm0.2$
($\chi^2_{\nu}=0.64$, probability = 74\%).

Fits of the radial profile of r3-16's optical counterpart candidate in
the HST images to the HST PSF (described in detail in
\citealp{dolphin2000}) yield $\chi > 3$ for all observations.  This
statistic is normalized to have a median value of 1 for single stars.
In addition, sharpness measurements for the object all 3 epochs yield
values $< -0.3$.  Tests of the HSTPHOT software suggest normal stars
have $\chi < 3$ and sharpness between -0.3 and 0.3
\citep{dolphin2000}.  These measurements suggest that the object is
extended; such $\chi$ and sharpness values are typical of unresolved
binaries.  On the PC chip of WFPC2, where we measure 7 point sources
to have a FWHM of 0.10$\pm$0.01$''$, r3-16 has a FWHM of
0.23$\pm$0.01$''$, implying an intrinsic size of 0.21$\pm$0.02$''$ or
0.82$\pm$0.08 parsec if located in M31.  An independent measurement on
the WF3 chip, where 10 point sources have a FWHM of 0.18$\pm$0.04$''$,
has a FWHM of 0.39$\pm$0.03$''$, implying an intrinsic size of
0.35$\pm$0.05$''$, about 50\% larger than the result from the PC.  The
width of the the same object in the LGS ground based $U$-band images,
where we measure 10 nearby point sources to have a FWHM of
1.15$\pm$0.15$''$, is 1.53$\pm$0.10$''$, implying an intrinsic width
of the object of 1.01$\pm$0.18$''$ or 3.9$\pm$0.7 parsec if located in
M31.

The discrepancies between the implied sizes of the object in
ground-based and HST images are difficult to reconcile.  The larger
size, like the brighter magnitude, measured in the ground-based images
may be due to crowding in the M31 bulge.  In any case, the object is
not likely to be a single star, though it may be a blended foreground
binary with a separation of $\lap$0.01 pc.  The X-ray spectrum,
optical size, brightness, variability and emission line properties of
r3-16 allow the possibility that it is a background active galactic
nucleus (AGN).  Such AGN have been seen through several other Local
Group galaxies at optical magnitudes similar to those measured for
r3-16 (e.g., \citealp{tinney1997}).  This possibility is also
consistent with the measured ${\rm N_H}$ which is higher than the
Galactic value to M31 of $7 \times 10^{20}$ cm$^{-2}$ , as the light
of a background AGN would be heavily absorbed by M31.  The increase in
apparent angular size with increasing pixel size would then be a
result of integrating more diffuse emission in the larger pixels.

Alternatively, r3-16 could be a CV at a distance of $\lap$1~kpc.  This
possibility is consistent with the optical brightness and colors.  The
X-ray luminosity would then be $\lap$10$^{32}$\ergs\/ during the
outburst, and $\lap$10$^{30}$\ergs\/ during quiescence.  This range is
typical of that seen in CVs (\citealp{patterson1985},
\citealp{warner1995}).  The $\sim$0.3 mag drop in the $U$-band along
with the drop in the X-ray flux does not argue for nor against the CV
hypothesis, as the expected relation between this relatively small
(for a CV outburst) $U$-band change and the X-ray flux is unclear.
The ACIS spectrum measured during outburst is harder than that
typically seen in CVs in outburst \citep{warner1995}, but there are
some CVs which show hard spectra during outburst \citep{silber1994}.
The ACIS spectrum also shows absorption above the value expected from
the Galaxy alone, which may be somewhat unusual but not unheard of for
CVs.  The 1~kpc distance should be taken as an upper limit because, at
the galactic latitude of M31 ($b_{II} = -21$), an object at that
distance would be 360~pc below the Galactic plane, which is at the
extreme range for CVs (\citealp{warner1995}, Section 9.5).  If r3-16
is a CV, the extended, persistent optical emission could be a nova
shell from an earlier (unseen) nova eruption of this CV.  Similar to
the images of r3-16, many nova remnants are only slightly extended in
ground-based and HST images \citep{gill2000, gill1998}.  A high
quality optical spectrum could uncover the true nature of this source.

\section{Conclusions}

We have combined 17 epochs of snapshot observations covering most of
the M31 disk with the {\it Chandra}-HRC.  These data have provided
detections of 166 discrete X-ray sources.  All but 7 of these have
been previously detected.  Comparison of the LF of the bulge sources
to that of the disk sources reveals significant differences in shape.
The slope of the disk LF is comparable to that of elliptical galaxies.
This similarity is consistent with the link between star formation
rate and LF slope, as the star formation rate in the M31 disk is
rather low.  Analysis of the spatial distribution of sources shows
that most of the brightest sources in the disk lie in the southwestern
half of the disk.

We have found candidate counterparts for 55 of the 166 sources at
longer wavelengths in previous surveys.  These counterparts come in a
variety of types, including SNRs, globular clusters, stars and an
extended optical source of unknown nature.  Fifteen stellar
counterpart candidates were detected in recent wide-field M31 data
taken by the Local Group Survey \citep{massey2001}.  Counterparts were
detected for 2 X-ray transients using data from HST. Analysis of one
counterpart (r2-67) found it to be an optical transient.  The
properties of this system are consistent with a BHXN in M31.  The
ratio of the optical to X-ray flux yields an estimate of the period of
this system of $23^{+54}_{-16}$ days.  This period is consistent with
the X-ray decline rate.  The optical properties of the other transient
(r3-16) are difficult to understand, but it may be a foreground blend,
a CV with an associated nova remnant, or a background AGN.

The long-term lightcurves of these sources suggest that at least 44 of
them varied significantly over the course of these observations, which
cover a baseline of about 2.5 years.  From the lightcurves, we have
selected 17 good transient candidates, and we have determined that at
any given time there are 2$\pm$1 active X-ray transients in M31.
The frequency of occurrence for these bright transient events suggests
that $\lap$1\% of the bright X-ray sources in the M31 bulge are new
transients.  If these sources are black hole containing LMXBs with
duty cycles of $\sim$1\%, the ratio of black hole to neutron star
primaries in LMXBs in M31 is $\sim$1, comparable to the ratio seen in
the Galaxy but greater than expectations from evolutionary
calculations.

Finally, it is unfortunate that we have been unsuccessful at finding
counterparts for more than 100 sources.  Many of these counterparts
are located in the extremely crowded bulge and will require very high
angular resolution optical data to recover.  These objects are likely
to be faint in the optical (V$\gap$20), and they may be heavily
absorbed.  Currently undetected counterparts in the disk are likely
even fainter or heavily absorbed.  Spectral X-ray data of the disk,
unavailable with the HRC data, will help to discern whether intrinsic
faintness or absorption are the cause for the lack of detection of
counterparts at longer wavelengths.

We thank Phil Kaaret for allowing us the use of his 50 ks observation
of the M31 nucleus to help constrain our lightcurves in that region
and for helping to measure the model fits.  We thank the LGS team
\citep{massey2001} for supplying the optical data to search for new
optical counterparts for the X-ray sources.  Support for this work was
provided by NASA through grant number GO-9087 from the Space Telescope
Science Institute and through grant number GO-3103X from the {\it
Chandra} X-ray Center.  MRG acknowledges support from NASA LTSA grant
NAG5-10889.  Finally we thank our referee, Phil Charles, for his many
helpful and insightful comments which greatly helped to improve this
paper.

\vspace{-0.5cm}

\clearpage
\begin{deluxetable}{cccccc}
\tablewidth{5.0in}
\tablecaption{Summary of {\it Chandra} HRC-I Observations}
\tableheadfrac{0.01}
\tablehead{
\colhead{\bf{OBSID}} &
\colhead{\bf{Date}} &
\colhead{\bf{Epoch}} &
\colhead{\bf{R.A.}} &
\colhead{\bf{DEC}} &
\colhead{\bf{Exposure (s)}} 
}
\startdata
\tablevspace{-0.01in}
243 & 1999-11-30  &  1  & 0:40:27.00 & 40:40:12.0 &  1163.613\\
255 & 1999-11-30  &  1  & 0:42:08.00 & 40:55:17.0 &  1269.114\\
267 & 1999-11-30  &  1  & 0:42:44.40 & 41:16:08.3 &  1270.082\\
279 & 1999-11-30  &  1  & 0:44:07.00 & 41:43:16.0 &  2683.351\\
291 & 1999-11-30  &  1  & 0:45:20.00 & 41:49:47.0 &  1270.183\\
268 & 1999-12-23  &  2  & 0:42:44.40 & 41:16:08.3 &  5178.063\\
245 & 2000-01-19  &  3  & 0:40:27.00 & 40:40:12.0 &  1159.204\\
257 & 2000-01-19  &  3  & 0:42:08.00 & 40:55:17.0 &  1207.168\\
269 & 2000-01-19  &  3  & 0:42:44.40 & 41:16:08.3 &  1205.910\\
281 & 2000-01-19  &  3  & 0:44:07.00 & 41:43:16.0 &  1205.035\\
293 & 2000-01-19  &  3  & 0:45:20.00 & 41:49:47.0 &  1208.001\\
246 & 2000-02-13  &  4  & 0:40:27.00 & 40:40:12.0 &  1417.755\\
258 & 2000-02-13  &  4  & 0:42:08.00 & 40:55:17.0 &  1474.374\\
270 & 2000-02-13  &  4  & 0:42:44.40 & 41:16:08.3 &  1469.659\\
282 & 2000-02-13  &  4  & 0:44:07.00 & 41:43:16.0 &  1473.856\\
294 & 2000-02-13  &  4  & 0:45:20.00 & 41:49:47.0 &  1467.671\\
247 & 2000-03-08  &  5  & 0:40:27.00 & 40:40:12.0 &  2162.940\\
259 & 2000-03-08  &  5  & 0:42:08.00 & 40:55:17.0 &  2472.312\\
271 & 2000-03-08  &  5  & 0:42:44.40 & 41:16:08.3 &  2461.785\\
283 & 2000-03-08  &  5  & 0:44:07.00 & 41:43:16.0 &  2476.503\\
295 & 2000-03-08  &  5  & 0:45:20.00 & 41:49:47.0 &  2455.864\\
248 & 2000-05-26  &  6  & 0:40:27.00 & 40:40:12.0 &  1156.204\\
260 & 2000-05-26  &  6  & 0:42:40.80 & 40:51:54.0 &  1209.234\\
272 & 2000-05-26  &  6  & 0:42:44.40 & 41:16:08.3 &  1202.670\\
284 & 2000-05-26  &  6  & 0:44:07.00 & 41:43:16.0 &  1211.161\\
296 & 2000-05-26  &  6  & 0:45:20.00 & 41:49:47.0 &  1204.171\\
249 & 2000-06-21  &  7  & 0:40:27.00 & 40:40:12.0 &  1189.396\\
261 & 2000-06-21  &  7  & 0:42:40.80 & 40:51:54.0 &  1186.803\\
273 & 2000-06-21  &  7  & 0:42:44.40 & 41:16:08.3 &  1190.727\\
285 & 2000-06-21  &  7  & 0:44:07.00 & 41:43:16.0 &  1186.797\\
297 & 2000-06-21  &  7  & 0:45:20.00 & 41:49:47.0 &  893.869\\
251 & 2000-08-18  &  8  & 0:40:27.00 & 40:40:12.0 &  977.143\\
263 & 2000-08-18  &  8  & 0:42:40.80 & 40:51:54.0 &  1184.931\\
275 & 2000-08-18  &  8  & 0:42:44.40 & 41:16:08.3 &  1188.885\\
287 & 2000-08-18  &  8  & 0:44:07.00 & 41:43:16.0 &  1186.985\\
299 & 2000-08-18  &  8  & 0:45:20.00 & 41:49:47.0 &  1170.571\\
252 & 2000-09-11  &  9  & 0:40:27.00 & 40:40:12.0 &  1189.792\\
264 & 2000-09-11  &  9  & 0:42:40.80 & 40:51:54.0 &  1189.081\\
276 & 2000-09-11  &  9  & 0:42:44.40 & 41:16:08.3 &  1182.794\\
288 & 2000-09-11  &  9  & 0:44:07.00 & 41:43:16.0 &  1193.216\\
300 & 2000-09-11  &  9  & 0:45:20.00 & 41:49:47.0 &  1150.291\\
253 & 2000-10-12  &  10  & 0:40:27.00 & 40:40:12.0 &  871.491\\
265 & 2000-10-12  &  10  & 0:42:40.80 & 40:51:54.0 &  1189.515\\
277 & 2000-10-12  &  10  & 0:42:44.40 & 41:16:08.3 &  1187.209\\
289 & 2000-10-12  &  10  & 0:44:07.00 & 41:43:16.0 &  1190.742\\
301 & 2000-10-12  &  10  & 0:45:20.00 & 41:49:47.0 &  1162.660\\
254 & 2000-11-17  &  11  & 0:40:27.00 & 40:40:12.0 &  1182.351\\
266 & 2000-11-17  &  11  & 0:42:40.80 & 40:51:54.0 &  1180.444\\
278 & 2000-11-17  &  11  & 0:42:44.40 & 41:16:08.3 &  1182.300\\
290 & 2000-11-17  &  11  & 0:44:07.00 & 41:43:16.0 &  1176.379\\
302 & 2000-11-17  &  11  & 0:45:20.00 & 41:49:47.0 &  979.562\\
1565 & 2001-01-31  &  12  & 0:40:27.00 & 40:40:12.0 &  964.220\\
1567 & 2001-02-01  &  12  & 0:42:08.00 & 40:55:17.0 &  1172.113\\
1569 & 2001-02-01  &  12  & 0:42:44.40 & 41:16:08.3 &  1171.771\\
1571 & 2001-02-01  &  12  & 0:44:07.00 & 41:43:16.0 &  1187.686\\
1573 & 2001-02-01  &  12  & 0:45:20.00 & 41:49:47.0 &  1186.826\\
1566 & 2001-06-10  &  13  & 0:40:27.00 & 40:40:12.0 &  871.495\\
1568 & 2001-06-10  &  13  & 0:42:08.00 & 40:55:17.0 &  1183.653\\
1570 & 2001-06-10  &  13  & 0:42:44.40 & 41:16:08.3 &  1184.236\\
1572 & 2001-06-10  &  13  & 0:44:07.00 & 41:43:16.0 &  1184.442\\
1574 & 2001-06-10  &  13  & 0:45:20.00 & 41:49:47.0 &  1194.797\\
2886 & 2001-09-08  &  14  & 0:40:27.10 & 40:40:12.0 &  867.664\\
2890 & 2001-09-08  &  14  & 0:42:07.90 & 40:55:15.6 &  1182.953\\
2903 & 2001-09-08  &  14  & 0:42:44.40 & 41:43:15.6 &  1188.012\\
2907 & 2001-09-08  &  14  & 0:44:07.00 & 41:43:15.6 &  1179.696\\
2911 & 2001-09-08  &  14  & 0:45:19.90 & 41:49:48.0 &  1178.814\\
2887 & 2001-11-19  &  15  & 0:40:27.10 & 40:40:12.0 &  982.551\\
2891 & 2001-11-19  &  15  & 0:42:07.90 & 40:55:15.6 &  1173.808\\
2904 & 2001-11-19  &  15  & 0:42:44.40 & 41:16:08.3 &  1181.646\\
2908 & 2001-11-19  &  15  & 0:44:07.00 & 41:43:15.6 &  1172.898\\
2912 & 2001-11-19  &   15  & 0:45:19.90 & 41:49:48.0 &  1161.823\\
2888 & 2002-01-16  &  16  & 0:40:27.10 & 40:40:12.0 &  975.687\\
2892 & 2002-01-16  &   16  & 0:42:07.90 & 40:55:15.6 &  1083.539\\
2905 & 2002-01-16  &   16  & 0:42:44.40 & 41:16:08.3 &  1094.222\\
2909 & 2002-01-16  &   16  & 0:44:07.00 & 41:43:15.6 &  1084.598\\
2913 & 2002-01-16  &   16  & 0:44:28.40 & 41:56:28.1 &  1097.045\\
2889 & 2002-06-02  &   17  & 0:40:27.10 & 40:40:12.0 &  980.472\\
2893 & 2002-06-02  &  17  & 0:42:07.90 & 40:55:15.6 &  1190.200\\
2906 & 2002-06-02  &   17  & 0:42:44.40 & 41:16:08.3 &  1187.762\\
2910 & 2002-06-02  &   17  & 0:44:07.00 & 41:43:15.6 &  1196.402\\
2914 & 2002-06-02  &   17  & 0:45:19.90 & 41:49:48.0 &  1190.132\\
\tablevspace{-0.05in}
\enddata
\label{obslist}
\end{deluxetable}

\clearpage
\begin{landscape}
\begin{deluxetable}{ccccccccr}
\tablewidth{8.8in}
\tablecaption{Luminosities, Variability, Other Detections, References and Object Types for the X-ray Sources}
\tableheadfrac{0.01}
\tablehead{
\colhead{\bf{OBJ$^a$}} &
\colhead{\bf{R.A.$^b$}} &
\colhead{\bf{DEC$^b$}} &
\colhead{\bf{cts$^c$}} &
\colhead{\bf{S/N$^d$}} &
\colhead{\bf{L$^e$}} &
\colhead{\bf{$\chi^2_{\nu}$ $^f$}} &
\colhead{\bf{Counterpart$^g$}} &
\colhead{\bf{Ref.$^h$ (type)$^i$}}
}
\startdata
s1-74 &  0:39:56.34 & 40:41:00.9 &  51 &     7.0 & 5.4$\pm$1.0 &   0.242 &  foreground (V$\lap$14.7)  &  new (*)\\
s1-75 &  0:40:13.77 & 40:50:05.1 & 2215 &   118.0 & 265.9$\pm$6.0 &   1.968 &  B3 0037+405  &  23 (BLL)\\
s1-76 &  0:40:14.34 & 40:33:41.5 &  65 &     9.3 & 7.1$\pm$1.1 &   0.263 &  new  &  new (X)\\
s1-77 &  0:40:20.29 & 40:43:58.5 & 1436 &   169.9 & 146.9$\pm$4.0 &   1.086 &  Bol 5  &  6 (GC)\\
s1-78 &  0:40:22.71 & 40:36:10.5 &  42 &     6.9 & 4.3$\pm$0.8 &   0.253 &  LGSJ004022.7+403610  &  new (*)\\
s2-63 &  0:40:24.20 & 40:29:48.5 &  32 &     5.2 & 3.8$\pm$0.9 &   0.356 &  RX J004024.5+402946  &  1 (X)\\
s1-79 &  0:40:55.16 & 40:56:03.5 & 850 &    27.4 & 90.9$\pm$4.5 &  18.805 &  new  &  new (X)\\
s1-45 &  0:41:18.55 & 40:52:00.5 & 136 &     8.2 & 9.2$\pm$1.3 &   1.251 &  foreground (V$\lap$14.7)  &  QSS (*)\\
s1-31 &  0:41:25.78 & 40:58:47.5 & 171 &    11.7 & 20.3$\pm$2.2 &   1.024 &  RX J004125.9+405842  &  1 (X)\\
s1-34 &  0:41:25.93 & 40:53:25.5 &  59 &     5.0 & 4.9$\pm$1.1 &   0.437 &  LGSJ004126.2+405327  &  new (*)\\
s1-28 &  0:41:37.60 & 41:01:09.8 &  39 &     3.8 & 4.1$\pm$1.2 &   0.285 &  RX J004138.3+410106  &  1 (X)\\
s1-16 &  0:41:49.81 & 41:01:05.8 &  20 &     3.7 & 1.7$\pm$0.5 &   0.146 &  RX J004149.8+410109  &  1 (X)\\
s1-80 &  0:41:54.63 & 40:56:47.4 & 150 &    24.0 & 13.9$\pm$1.2 &   7.158 &  1RXS J004154.1+405648 (ATel97)  &  16 (X)\\
s1-64 &  0:42:02.92 & 40:46:11.4 & 137 &     9.3 & 12.9$\pm$1.7 &   0.695 &  LGSJ004203.0+404613  &  new (*)\\
r3-125 &  0:42:05.71 & 41:13:29.7 &  65 &     5.3 & 5.2$\pm$1.1 &   0.620 &  [PFJ93] 3  &  2 (X)\\
s1-12 &  0:42:06.12 & 41:02:47.5 &  78 &     5.6 & 4.2$\pm$0.8 &   0.469 &  Bol D42  &  6 (GC)\\
s1-11 &  0:42:07.16 & 41:00:17.9 &  50 &     5.8 & 3.8$\pm$0.8 &   0.273 &  Bol D44  &  6 (GC)\\
r3-61 &  0:42:07.74 & 41:18:15.4 & 338 &    34.2 & 26.6$\pm$1.6 &   0.953 &  C84 10  &  3 (X)\\
s1-9 &  0:42:07.89 & 41:04:34.6 &  82 &     5.3 & 4.2$\pm$0.9 &   0.408 &  RX J004208.2+410438  &  1 (X)\\
r3-60 &  0:42:09.11 & 41:20:49.0 &  71 &     6.8 & 5.7$\pm$1.0 &   0.547 &  CXOM31 J004209.0+412048  &  15 (X)\\
r3-59 &  0:42:09.53 & 41:17:45.3 &  80 &     9.6 & 6.2$\pm$0.9 &   0.350 &  mita140  &  14 (GC)\\
r3-58 &  0:42:10.32 & 41:15:10.1 &  52 &     7.3 & 4.0$\pm$0.7 &   0.795 &  [PFJ93] 6  &  2 (X)\\
r3-57 &  0:42:11.01 & 41:12:47.7 &  34 &     4.9 & 2.7$\pm$0.7 &   0.139 &  CXOM31 J004210.9+411248  &  15 (X)\\
r3-56 &  0:42:11.79 & 41:10:48.6 &  61 &     5.7 & 4.9$\pm$1.0 &   0.285 &  [PFJ93] 7  &  2 (X)\\
r3-55 &  0:42:11.92 & 41:16:48.9 &  32 &     5.5 & 2.5$\pm$0.6 &   0.177 &  CXOM31J004211.9+411648  &  15 (X)\\
s1-58 &  0:42:12.12 & 40:53:35.6 &  25 &     4.7 & 2.2$\pm$0.6 &   0.212 &  MLA 462   &  17 (X)\\
r3-54 &  0:42:12.15 & 41:17:59.1 & 140 &    17.8 & 10.8$\pm$1.1 &   1.704 &  mita153  &  14 (GC)\\
r3-52 &  0:42:13.14 & 41:18:36.7 & 647 &    60.4 & 49.9$\pm$2.1 &   4.630 &  C84 13  &  3 (X)\\
s1-81 &  0:42:13.77 & 40:39:25.0 &  83 &     4.7 & 11.8$\pm$2.7 &   0.665 &  RX J004213.5+403917  &  1 (X)\\
r3-50 &  0:42:15.13 & 41:12:34.6 & 342 &    37.7 & 26.4$\pm$1.6 &   3.871 &  [PFJ93] 10  &  2 (X)\\
r3-48 &  0:42:15.37 & 41:20:31.7 &  77 &     9.9 & 6.0$\pm$0.9 &   0.206 &  [PFJ93] 11  &  2 (X)\\
r3-47 &  0:42:15.65 & 41:17:21.0 & 149 &    19.5 & 11.2$\pm$1.1 &   0.988 &  C84 16  &  3 (X) \\
s1-7 &  0:42:15.86 & 41:01:14.7 & 1214 &    66.6 & 74.7$\pm$2.4 &   2.675 &  mita159  &  14 (GC)\\
s1-6 &  0:42:16.55 & 40:55:52.3 & 199 &    25.8 & 17.5$\pm$1.4 &   0.442 &  RX J004216.4+405555  &  1 (X)\\
r3-46 &  0:42:17.06 & 41:15:08.8 &  43 &     9.0 & 3.2$\pm$0.6 &   5.710 &  CXOM31 J004216.9+411508  &  15 (X)\\
r3-45 &  0:42:18.32 & 41:12:24.1 & 643 &    65.8 & 49.2$\pm$2.0 &   1.398 &  C84 21  &  3 (X) \\
r3-44 &  0:42:18.62 & 41:14:01.9 & 574 &    77.8 & 43.0$\pm$1.9 &   1.307 &  mita164  &  14 (GC)\\
s1-82 &  0:42:20.82 & 40:51:36.3 &  27 &     5.0 & 2.4$\pm$0.6 &   0.552 &  new  &  new (X)\\
r3-42 &  0:42:21.48 & 41:16:01.6 & 406 &    66.6 & 29.8$\pm$1.5 &   1.350 &  C84 23  &  3 (X)\\
r3-41 &  0:42:21.54 & 41:14:18.5 &  46 &     8.4 & 3.4$\pm$0.6 &   0.557 &  CXOM31 J004221.5+411419  &  15 (X)\\
s1-5 &  0:42:22.20 & 40:59:25.6 & 234 &    24.2 & 16.4$\pm$1.2 &   0.448 &  RX J004222.1+405926  &  1  (X)\\
r3-40 &  0:42:22.41 & 41:13:34.2 & 293 &    44.4 & 21.8$\pm$1.3 &   1.429 &  C84 25  &  3 (X)\\
r3-39 &  0:42:22.94 & 41:15:35.4 & 884 &   125.9 & 64.6$\pm$2.2 &  44.079 &  C84 26  &  3 (X)\\
r3-38 &  0:42:23.15 & 41:14:07.4 & 141 &    26.8 & 10.4$\pm$0.9 &   1.193 &  C84 27  &  3 (X)\\
r2-45 &  0:42:25.11 & 41:13:40.7 & 109 &    19.7 & 8.0$\pm$0.8 &   1.287 &  C84 28  &  3 (X)\\
r3-87 &  0:42:25.72 & 41:25:50.3 &  81 &     9.7 & 7.0$\pm$1.0 &   0.155 &  [PFJ93] 24  &  2 (X)\\
r2-36 &  0:42:26.03 & 41:19:15.3 & 155 &    26.7 & 11.4$\pm$1.0 &   0.962 &  mita174  &  14 (GC)\\
r3-36 &  0:42:28.20 & 41:10:00.7 & 310 &    34.2 & 24.2$\pm$1.5 &   0.638 &  CXOM31 J004228.1+410959  &  15 (X)\\
r2-35 &  0:42:28.28 & 41:12:23.2 & 816 &   113.7 & 60.7$\pm$2.2 &   1.878 &  C84 30  &  3 (X)\\
r3-111 &  0:42:28.92 & 41:04:35.8 & 608 &    23.3 & 29.9$\pm$1.7 &   2.483 &  C84 31  &  3 (X)\\
r2-34 &  0:42:31.13 & 41:16:21.9 & 734 &   133.3 & 52.1$\pm$1.9 &   0.526 &  C84 36  &  3 (X)\\
r2-33 &  0:42:31.24 & 41:19:39.2 & 156 &    30.8 & 11.5$\pm$1.0 &   2.727 &  mita192  &  14 (GC)\\
r2-32 &  0:42:32.07 & 41:13:14.5 & 411 &    71.9 & 29.8$\pm$1.5 &   3.280 &  C84 39  &  3 (X)\\
r2-68 &  0:42:32.63 & 41:17:02.9 &  19 &     4.6 & 1.4$\pm$0.4 &   0.149 &  new  &  new (X)\\
r2-31 &  0:42:32.74 & 41:13:11.0 &  65 &    13.6 & 4.7$\pm$0.6 &   0.801 &  RX J004232.7+411318  &  1 (X) \\
r2-30 &  0:42:33.87 & 41:16:20.2 & 106 &    21.4 & 7.5$\pm$0.8 &   0.264 &  [PFJ93] 32  &  2 (X)\\
r2-29 &  0:42:34.43 & 41:18:09.6 &  40 &     9.5 & 2.8$\pm$0.5 &   1.149 &  CXOM31 J004234.4+411809  &  15 (X)\\ 
n1-74 &  0:42:34.53 & 41:32:47.7 & 159 &     5.5 & 23.4$\pm$4.5 &   1.207 &  Bol 116  &  6 (GC)\\
r2-28 &  0:42:34.75 & 41:15:23.5 &  69 &    14.6 & 4.8$\pm$0.6 &   2.119 &  CXOM31 J004234.7+411523  &  15 (X)\\
s1-1 &  0:42:34.97 & 40:57:21.5 &  79 &    10.5 & 6.5$\pm$0.9 &   1.399 &  new  &  new (X)\\
s1-83 &  0:42:35.03 & 40:48:38.0 & 160 &    14.6 & 15.4$\pm$1.6 &   0.866 &  Bol D63  &  6 (GC)\\
r2-27 &  0:42:35.18 & 41:20:05.8 & 114 &    23.9 & 8.4$\pm$0.8 &   0.567 &  [PFJ93] 34  &  2 (X)\\
r2-42 &  0:42:36.64 & 41:13:50.4 &  31 &     7.1 & 2.2$\pm$0.5 &   0.154 &  QSS  &  QSS (X)\\
r2-26 &  0:42:38.57 & 41:16:03.9 & 5579 &   589.9 & 390.2$\pm$5.3 &   7.392 &  [PFJ93] 35  &  2 (X)\\
r2-25 &  0:42:39.52 & 41:14:28.7 &  97 &    22.1 & 6.9$\pm$0.7 &   0.999 &  TF 42  &  3 (X)\\
r1-34 &  0:42:39.58 & 41:16:14.6 &  26 &     5.9 & 1.9$\pm$0.4 &  42.880 &  TF 43  &  3 (X)\\
s1-84 &  0:42:39.70 & 40:43:18.7 & 114 &     5.8 & 12.7$\pm$2.4 &   0.359 &  RX J004238.8+404317  &  1 (X)\\
r1-15 &  0:42:39.98 & 41:15:47.7 & 424 &    66.7 & 29.6$\pm$1.5 &   2.101 &  Ford M 31 17  &  5 (PN)\\
r2-24 &  0:42:40.20 & 41:18:45.6 &  74 &    17.5 & 5.3$\pm$0.7 &   0.431 &  CXOM31 J004240.1+411845  &  15 (X)\\
r2-22 &  0:42:40.64 & 41:13:27.5 & 173 &    33.7 & 12.3$\pm$1.0 &   2.230 &  [PFJ93] 38  &  2 (X)\\
r3-34 &  0:42:40.65 & 41:10:32.8 &  26 &     4.5 & 2.0$\pm$0.5 &   0.296 &  mita212  &  14 (GC)\\
s1-85 &  0:42:40.66 & 40:51:17.5 & 161 &    25.8 & 15.0$\pm$1.3 &   7.419 &   M32-t1  &  22 (X)\\
r1-32 &  0:42:41.43 & 41:15:24.4 &  50 &    10.9 & 3.5$\pm$0.6 &   1.328 &  mita213  &  14 (GC)\\
r3-31 &  0:42:41.65 & 41:21:06.7 &  30 &     4.5 & 2.2$\pm$0.6 &   0.408 &  CXOM31 J004241.6+412106  &  15 (X)\\ 
r1-31 &  0:42:42.07 & 41:15:32.4 &  23 &     5.4 & 1.6$\pm$0.4 &   0.164 &  CXOM31 J004242.0+411532  &  15 (X)\\
r2-69 &  0:42:42.16 & 41:14:24.4 &  28 &     6.0 & 2.0$\pm$0.5 &   0.830 &  new  &  new (X)\\
r1-5 &  0:42:42.16 & 41:16:08.4 & 940 &   129.2 & 65.5$\pm$2.2 &  46.434 &  CXOM31 J004242.1+411608  &  15 (X)\\
r2-21 &  0:42:42.33 & 41:14:45.6 & 180 &    38.0 & 12.6$\pm$1.0 &   0.256 &  [PFJ93] 39  &  2 (X)\\
s1-32 &  0:42:42.45 & 40:51:53.0 & 844 &    71.8 & 78.5$\pm$2.9 &   1.091 &  bbk2000-7442 (M32)  &  7 (*)\\
r1-14 &  0:42:42.47 & 41:15:53.8 & 224 &    35.0 & 15.6$\pm$1.1 &   0.490 &  [PFJ93] 41  &  2 (X)\\
r1-30 &  0:42:42.60 & 41:16:59.8 &  31 &     6.1 & 2.2$\pm$0.5 &   0.136 &  [PFJ93] 40  &  2 (X)\\
r1-13 &  0:42:42.98 & 41:15:43.3 & 199 &    42.3 & 13.9$\pm$1.0 &   0.592 &  [PFJ93] 42  &  2 (X)\\
r1-35 &  0:42:43.12 & 41:16:04.1 &  39 &     8.2 & 2.7$\pm$0.5 &   0.817 &  SSS near SN 1885A  &  18 (SN)\\
r1-24 &  0:42:43.29 & 41:16:40.1 &  36 &     7.0 & 2.5$\pm$0.5 &   0.215 &  Ford M 31 322  &  5 (PN)\\
r2-19 &  0:42:43.30 & 41:13:19.7 &  92 &    20.4 & 6.6$\pm$0.7 &   0.726 &  QSS  &  QSS (X)\\
r1-12 &  0:42:43.73 & 41:16:32.7 & 170 &    30.7 & 11.9$\pm$1.0 &   1.574 &  [PFJ93] 43  &  2 (X)\\
r1-28 &  0:42:43.77 & 41:15:14.2 &  40 &     8.3 & 2.8$\pm$0.5 &   1.195 &  CXOM31 J004243.7+411514  &  15 (X)\\
r1-23 &  0:42:43.84 & 41:16:04.1 &  45 &     8.2 & 3.2$\pm$0.6 &   0.566 &  Ford M 31 8  &  5 (PN)\\
r1-11 &  0:42:43.86 & 41:16:29.8 &  88 &    18.6 & 6.1$\pm$0.7 &   0.707 &  [PFJ93] 43  &  2 (X)\\
r1-22 &  0:42:44.22 & 41:16:14.7 &  43 &     6.9 & 3.0$\pm$0.6 &   0.230 &  BFS98-M31-53  &  8 (*)\\
r1-9 &  0:42:44.36 & 41:16:07.5 & 470 &    62.2 & 32.7$\pm$1.6 &  8.308 &  SSS  &  SSS (X)\\
r3-30 &  0:42:44.39 & 41:11:58.7 &  39 &     9.2 & 2.9$\pm$0.5 &   0.289 &  CXOM31 J004244.4+411157  &  15 (X)\\
r1-8 &  0:42:44.66 & 41:16:18.7 &  80 &    13.2 & 5.6$\pm$0.7 &   0.458 &  [PFJ93] 47  &  2 (X)\\
r3-29 &  0:42:44.84 & 41:11:38.2 & 412 &    62.4 & 30.6$\pm$1.6 &   2.189 &  C84 67  &  3 (X)\\
r2-18 &  0:42:44.89 & 41:17:40.2 &  66 &    13.3 & 4.6$\pm$0.6 &   0.505 &  CXOM31 J004244.8+411739  &  15 (X)\\
r2-17 &  0:42:45.09 & 41:14:07.4 &  36 &     9.8 & 2.6$\pm$0.5 &   0.182 &  [PFJ93] 46  &  2 (X)\\
r1-4 &  0:42:45.10 & 41:16:21.9 & 331 &    49.9 & 23.1$\pm$1.3 &   4.790 &  [PFJ93] 47  &  2 (X)\\
r1-26 &  0:42:45.11 & 41:15:23.4 &  44 &     9.7 & 3.1$\pm$0.5 &   0.304 &  Ford M 31 21  &  5 (PN)\\
r2-16 &  0:42:45.19 & 41:17:22.5 & 111 &    24.0 & 7.8$\pm$0.8 &   2.575 &  CXOM31 J004245.2+411722  &  15 (X)\\
r1-20 &  0:42:45.20 & 41:16:11.2 &  56 &    10.2 & 3.9$\pm$0.6 &   0.160 &  BFS98-M31-419  &  8 (*)\\
r1-7 &  0:42:45.59 & 41:16:08.5 &  52 &     9.5 & 3.6$\pm$0.6 &   0.395 &  CXOM31 J004245.5+411608  &  15 (X)\\
r1-19 &  0:42:46.00 & 41:16:19.9 &  37 &     8.0 & 2.6$\pm$0.5 &   0.518 &  CXOM31 J004245.9+411619  &  15 (X)\\
r1-18 &  0:42:46.18 & 41:15:43.1 &  29 &     6.7 & 2.0$\pm$0.4 &   0.208 &  CXOM31 J004246.1+411543  &  15 (X)\\
r3-28 &  0:42:46.95 & 41:21:21.5 &  32 &     5.1 & 2.4$\pm$0.6 &   0.213 &  LGSJ004246.9+412120  &  new (*)\\
r1-3 &  0:42:46.97 & 41:16:15.9 & 318 &    42.5 & 22.2$\pm$1.3 &  2.157 &  [PFJ93] 49  &  2 (X)\\
r1-2 &  0:42:47.16 & 41:16:28.6 & 1004 &   150.4 & 70.2$\pm$2.2 &   8.826 &  Ford M 31 13  &  5 (PN)\\
r3-27 &  0:42:47.27 & 41:11:57.6 &  33 &     7.8 & 2.4$\pm$0.5 &   0.210 &  CXOM31J004247.2+411157  &  15 (X)\\
r1-17 &  0:42:47.84 & 41:16:23.0 &  30 &     6.9 & 2.1$\pm$0.4 &   0.327 &  CXOM31J004247.8+411623  &  15 (X)\\
r1-6 &  0:42:47.88 & 41:15:33.2 & 348 &    57.1 & 24.4$\pm$1.4 &   3.205 &  [PFJ93] 52  &  2 (X)\\
r1-25 &  0:42:47.90 & 41:15:49.8 &  18 &     4.7 & 1.2$\pm$0.3 &   0.180 &  SSS  &  SSS (X)\\
r1-1 &  0:42:48.52 & 41:15:21.4 & 1089 &   175.2 & 76.5$\pm$2.3 &   1.527 &  [PFJ93] 54  &  2 (X)\\
r3-25 &  0:42:48.55 & 41:25:22.7 & 179 &    16.6 & 15.0$\pm$1.4 &   2.768 &  RX J004248.7+412522  &  1 (X)\\
r1-16 &  0:42:48.69 & 41:16:24.9 &  46 &     8.9 & 3.2$\pm$0.6 &   0.417 &  CXOM31 J004248.7+411624  &  15 (X)\\
r2-14 &  0:42:49.20 & 41:18:15.9 &  84 &    19.6 & 6.0$\pm$0.7 &   0.268 &  [PFJ93] 55  &  2 (X)\\
n1-75 &  0:42:51.98 & 41:31:07.7 & 2436 &    54.1 & 212.5$\pm$5.8 &   1.233 &  mita225  &  14 (GC)\\
r2-13 &  0:42:52.51 & 41:18:54.8 & 1029 &   165.4 & 74.5$\pm$2.4 &   1.071 &  [PFJ93] 57  &  2 (X)\\
r2-12 &  0:42:52.52 & 41:15:40.2 & 928 &   159.9 & 65.5$\pm$2.2 &   3.345 &  SSS  &  SSS (X)\\
r3-69 &  0:42:53.53 & 41:25:55.3 &  45 &     5.7 & 3.8$\pm$0.8 &   0.197 &  DDB 1-13  &  11 (SNR)\\
r2-11 &  0:42:54.92 & 41:16:03.5 & 973 &   169.3 & 69.1$\pm$2.2 &   0.650 &  [PFJ93] 60  &  2 (X)\\
r2-10 &  0:42:55.14 & 41:18:36.7 &  68 &    16.1 & 4.9$\pm$0.6 &   0.357 &  CXOM31 J004255.1+411836  &  15 (X)\\
r3-23 &  0:42:55.25 & 41:25:55.2 & 102 &    12.0 & 8.7$\pm$1.1 &   0.304 & [PFJ93] 61   &  2 (X)\\
r2-9 &  0:42:55.58 & 41:18:35.3 &  48 &    11.9 & 3.5$\pm$0.6 &   0.352 &  Bol 138  &  6 (GC)\\
r2-8 &  0:42:56.89 & 41:18:44.0 &  42 &    10.4 & 3.1$\pm$0.5 &   1.121 &  LGSJ004257.0+411844  &  new (*)\\
r3-22 &  0:42:57.90 & 41:11:04.9 & 414 &    51.1 & 31.6$\pm$1.6 &   1.638 &  RX J004257.7+411103  &  1 (X)\\
r2-7 &  0:42:58.33 & 41:15:29.2 & 155 &    31.6 & 11.2$\pm$0.9 &   1.431 &  [PFJ93] 65  &  2 (X)\\
r2-63 &  0:42:59.28 & 41:16:43.5 &  43 &    10.3 & 3.1$\pm$0.5 &   0.689 &  SSS  &  SSS (X)\\
r2-6 &  0:42:59.64 & 41:19:19.7 & 490 &    79.9 & 36.4$\pm$1.7 &   0.664 &  Bol 143  &  6 (GC)\\
r2-5 &  0:42:59.86 & 41:16:05.9 & 394 &    78.4 & 28.5$\pm$1.5 &   3.506 &  Bol 144  &  6 (GC)\\
n1-76 &  0:43:00.87 & 41:30:13.3 & 423 &     8.7 & 33.9$\pm$4.2 &   0.350 &  [PFJ93] 69  &  2 (X)\\
r2-37 &  0:43:01.07 & 41:13:51.4 &  30 &     7.0 & 2.2$\pm$0.5 &   0.256 &  [PFJ93] 68  &  2 (X)\\
r2-4 &  0:43:02.93 & 41:15:22.9 & 412 &    71.6 & 30.2$\pm$1.5 &   5.895 &  Bol 146  &  6 (GC)\\
r3-20 &  0:43:03.08 & 41:10:16.1 &  79 &    10.0 & 6.2$\pm$0.9 &   0.408 &  [PFJ93] 71  &  2 (X)\\
r2-3 &  0:43:03.23 & 41:15:28.2 & 345 &    62.8 & 25.3$\pm$1.4 &   6.468 &  CXOM31 J004303.2+411528  &  15 (X)\\
r3-19 &  0:43:03.34 & 41:21:22.0 & 198 &    24.3 & 15.3$\pm$1.2 &   0.217 &  mita240   &  14 (GC)\\
r2-2 &  0:43:03.85 & 41:18:05.3 & 314 &    54.1 & 23.3$\pm$1.4 &   3.745 &  Bol 148  &  6 (GC)\\
r2-1 &  0:43:04.25 & 41:16:01.4 &  59 &    12.6 & 4.3$\pm$0.6 &   0.544 &   CXOM31 J004304.2+411601  &  15 (X)\\
r2-67 &  0:43:05.66 & 41:17:03.3 & 222 &    43.0 & 16.5$\pm$1.1 & 517.664 &  LGSJ004305.6+411703  &  new (*)\\
r3-67 &  0:43:06.66 & 41:19:16.2 &  27 &     4.6 & 2.1$\pm$0.5 &   0.110 &   CXOM31 J004306.8+411912  &  15 (X)\\
r3-115 &  0:43:07.14 & 41:18:09.5 &  20 &     4.8 & 1.5$\pm$0.4 &   1.010 &  LGSJ004306.9+411809  &  SSS (*)\\
r3-18 &  0:43:07.52 & 41:20:19.8 &  49 &     7.5 & 3.7$\pm$0.7 &   0.528 &  mita246  &  14 (GC)\\
r3-17 &  0:43:08.63 & 41:12:50.1 &  81 &    12.1 & 6.2$\pm$0.8 &   0.312 &  [PFJ93] 74  &  2 (X)\\
r3-16 &  0:43:09.78 & 41:19:01.2 &  92 &    13.2 & 7.0$\pm$0.9 &  15.871 &  [WSB85] S1 4  &   20 (?) \\
r3-15 &  0:43:10.62 & 41:14:51.7 & 1248 &   150.5 & 94.3$\pm$2.7 &   2.681 &  mita251  &  14 (GC)\\
r3-13 &  0:43:13.16 & 41:18:14.1 &  28 &     5.1 & 2.1$\pm$0.5 &   0.187 &  LGSJ004318.1+411814  &  new (*)\\
r3-112 & 0:43:14.37 & 41:07:21.8 & 711 &    15.5 & 48.3$\pm$3.5 &   2.852 &  mita257  &  14 (GC)\\
r3-9 &  0:43:16.13 & 41:18:41.4 &  56 &     9.2 & 4.4$\pm$0.7 &   0.363 &  CXOM31 J004316.1+411841  &  15 (X)\\
r3-8 &  0:43:18.83 & 41:20:17.4 &  67 &     6.9 & 5.3$\pm$0.9 &   1.149 &  C84 101  &  3 (X)\\
r3-126 &  0:43:19.52 & 41:17:56.7 &  98 &    11.3 & 7.7$\pm$1.0 &   2.628 &  XMMU J004319.4+411759  &  19 (X)\\
r3-7 &  0:43:21.02 & 41:17:50.8 & 105 &    13.3 & 8.4$\pm$1.0 &   0.925 &  Ford M 31 209  &  5 (PN)\\
r3-63 &  0:43:27.95 & 41:18:31.0 & 199 &    12.8 & 16.4$\pm$1.7 &   0.196 &  DDB 1-15  &  11 (SNR)\\
r3-103 &  0:43:29.28 & 41:07:49.8 & 250 &    11.1 & 23.0$\pm$2.5 &   0.223 &  [PFJ93] 81  &  2 (X)\\
r3-3 &  0:43:32.50 & 41:10:38.9 &  53 &     7.4 & 4.6$\pm$0.8 &   0.356 &  RX J004331.9+411038  &  1 (X)\\
r3-2 &  0:43:34.31 & 41:13:26.6 & 262 &    23.8 & 22.5$\pm$1.6 &   2.120 &  C84 105  &  3 (X)\\
r3-1 &  0:43:37.27 & 41:14:43.8 & 405 &    32.3 & 35.0$\pm$2.0 &   0.355 &  mita299  &  14 (GC)\\
n1-77 &  0:43:44.56 & 41:24:17.1 & 281 &     8.1 & 24.7$\pm$3.3 &   0.954 &  2E 0040.9+4108  &  3 (X)\\
n1-78 &  0:43:45.64 & 41:36:57.4 &  47 &     6.8 & 4.2$\pm$0.8 &   0.570 &  Bol 193  &  6 (GC)\\
n1-79 &  0:43:53.56 & 41:16:51.0 & 200 &    19.1 & 18.9$\pm$1.6 &   0.448 &  D31J004353.8+411655.9  &  13 (*)\\
n1-80 &  0:44:02.73 & 41:39:26.0 &  31 &     4.9 & 2.2$\pm$0.6 &   0.378 &  RX J004402.4+413926  &  1 (X)\\
n1-81 &  0:44:22.64 & 41:45:06.7 & 134 &    12.6 & 6.5$\pm$0.7 &   0.804 &  LGSJ004422.6+414507  &  new (*)\\
n1-82 &  0:44:25.56 & 41:36:35.2 &  27 &     4.3 & 1.9$\pm$0.5 &   0.774 &  foreground (V$\lap$14.8)  &  new (*)\\
n1-83 &  0:44:38.02 & 41:45:14.4 & 132 &    11.1 & 6.3$\pm$0.8 &   0.766 &  [B90] 265  &  9 (Radio)\\
n1-84 &  0:44:48.88 & 41:47:27.9 &  71 &     4.8 & 3.5$\pm$0.8 &   0.142 &  new  &  new (X)\\
n1-61 &  0:45:11.18 & 41:45:56.4 &  97 &     4.7 & 4.8$\pm$1.1 &   0.282 &  RX J004510.9+414557  &  1 (X)\\
n1-85 &  0:45:28.05 & 41:54:10.9 & 101 &     4.1 & 7.9$\pm$2.0 &   2.281 &  SNR90  &  12 (SNR)\\
n1-17 &  0:45:45.61 & 41:39:41.6 & 5289 &   118.4 & 496.7$\pm$8.0 &  11.743 &  Bol 375  &  6 (GC)\\ 
n1-59 &  0:45:45.76 & 41:50:30.1 &  54 &     8.6 & 5.0$\pm$0.8 &   0.681 &  LGSJ004545.9+415030  &  new (*)\\
\enddata
\label{slist}
\end{deluxetable}
\end{landscape}
\newpage
\clearpage
{\footnotesize{ 

$^a$ Column 1 provides the object name that we use to reference the
object in this paper.  Identical sources have the same names in
\citet{kong2002} and \citet{distefano2003}.  The prefix r1 designates
objects in a square region 2$'$ $\times$ 2$'$ centered on the nucleus;
r2 designates objects outside of r1 but within a square region 8$'$
$\times$ 8$'$ centered on the nucleus; r3 designates objects outside
of r2 but within a square region 23$'$ $\times$ 23$'$ centered on the
nucleus; n1 designates objects outside of r3, north of the nucleus but
south of DEC = 42:01:00 (J2000); n2 designates objects north of DEC =
42:01:00 (J2000); s1 designates objects outside of r3, south of the
nucleus but north of DEC = 40:31:22 (J2000); s2 designates objects
south of DEC = 40:31:22 (J2000).  New objects within these regions are
numbered consecutively starting from the highest published number in
either \citet{kong2002} or \citet{distefano2003}.  The IAU sanctioned
names for the sources can be formulated by applying the RA and DEC to
the prefix CXOM31.  For example, the proper name of s1-74 is
CXOM31~J003956.3+404100.

$^b$ Columns 2 and 3 provide the Right Ascension and
Declination of the sources in J2000 coordinates.

$^c$ Column 4 shows the total net counts in the detection
of the source.

$^d$ Column 5 gives the $S/N$ measured by {\it wavdetect}.

$^e$ Column 6 provides the mean luminosity for the source
over all epochs in units of 10$^{36}\ erg\ s^{-1}$. This luminosity was
calculated by multiplying the measured flux by the conversion factor
discussed in \S 2.3.

$^f$ Column 7 lists the $\chi^2_{\nu}$ resulting from a fit
of the measured counting rate to a constant source at the mean rate.
A value greater than 1.47 indicates that the source has a 90\% chance
of being variable.

$^g$ Column 8 gives counterparts found at other wavelengths and in
previous X-ray surveys.  If the object was detected in previous X-ray
surveys as well as in other wavelengths, the optical (or, in one case,
radio) counterpart is listed.  If the object has no known counterpart
outside of the X-ray band, the earliest detection is listed as the
X-ray counterpart.  With the exception of the ``mita'' prefix,
counterpart names are IAU names taken from the Simbad database
(http://simbad.u-strasbg.fr/).  X-ray sources with no previous
detections and no optical counterparts are labeled ``new''.

$^h$ Column 9 provides the reference to the object
listed in Column 8, followed by an abbreviation in parentheses.  The
reference codes are as follows:  
1: \citet{supper2001};
2: \citet{primini1993};
3: \citet{trinchieri1991};
4: \citet{walterbos1992};
5: \citet{ford1978};
6: \citet{battistini1987};
7: \citet{brown2000};
8: \citet{brown1998};
9: \citet{braun1990};
11: \citet{dodorico1980};
12: \citet{williams1995};
13: \citet{mochejska2001};
14: \citet{magnier1993t};
15: \citet{kong2002};
16: \citet{voges1999};
17: \citet{meyssonnier1993};
18: \citet{devaucouleurs1985};
19: \citet{osborne2001};
20: \citet{wirth1985};
21: \citet{kaaret2002};
22: \citet{iauc7498};
23: \citet{perlman1996};
SSS,QSS: These are supersoft sources and quasisoft sources in the catalog of \citet{distefano2003};
new: this work.  New X-ray sources (labeled with an (X)) were seen for
the first time in this data set.  New optical counterparts (labeled
with a (*)) were found by searching the Local Group Survey data on the
M31 disk \citep{massey2001}, or they are new X-ray sources that have
no optical counterparts.

\hangindent 0.5cm
$^i$ The abbreviations in parentheses indicate the
following object types:\\
     (*) The source has a stellar counterpart.\\
     (X) The source has no optical counterpart, but a previous X-ray
detection is listed.\\
     (GC) The X-ray source lies in a known M31 globular cluster.\\
     (SNR) The X-ray source position is coincident with a known M31
supernova remnant.\\
     (PN) The X-ray source position is coincident with a known, unconfirmed
M31 planetary nebula.  These may be misidentified SNR.\\
     (BLL) The X-ray source position is coincident with a BL Lac candidate.\\
     (Radio) The source has also been detected in the radio.\\
     (?) The source has a counterpart of unknown nature.}}

\begin{deluxetable}{ccccc}
\tablewidth{0pt}
\tablecaption{Results of Broken Power-law Fits to the CLFs of Several Object Samples}
\tableheadfrac{0.01}
\tablehead{
\colhead{\bf{Sample}} &
\colhead{\bf{$\alpha_1$}\tablenotemark{a}} &
\colhead{\bf{$L_{break}$ (\ergs)}} &
\colhead{\bf{$\alpha_2$}\tablenotemark{b}} &
\colhead{\bf{Confidence}\tablenotemark{c}} 
}
\tablenotetext{a}{The slope of the CLF below the break luminosity.}
\tablenotetext{b}{The slope of the CLF above the break luminosity.}
\tablenotetext{c}{The fraction of Monte Carlo tests with fits statistically worse than the fit to our sample.}
\startdata
Disk & 0.6$\pm$0.3 & 2.6$^{+2.5}_{-0.9} \times 10^{37}$  & 1.5$^{+0.5}_{-0.4}$ & 0.44\\
Bulge (w/o GCs) &  0.5$\pm$0.2 & 7.0$^{+2.7}_{-1.3} \times 10^{37}$ & 1.7$^{+1.0}_{-0.6}$ & 0.50\\
Bulge (w/ GCs) & 0.5$\pm$0.2 & 7.1$^{+1.9}_{-1.5} \times 10^{37}$ & 1.9$^{+1.0}_{-0.6}$ & 0.47\\
Bulge (w/o Transients) & 0.4$\pm$0.2 & 7.0$^{+2.2}_{-1.3} \times 10^{37}$ & 1.8$^{+1.0}_{-0.6}$ & 0.35\\
\enddata
\label{bplfits}
\end{deluxetable}

\begin{deluxetable}{ccccc}
\tablewidth{0pt}
\tablecaption{Results of Power-law Fits to the CLFs of Several Object Samples}
\tableheadfrac{0.01}
\tablehead{
\colhead{\bf{Sample}} &
\colhead{\bf{$L\tablenotemark{a}_{min}$ (\ergs)}} &
\colhead{\bf{$\alpha$}} &
\colhead{\bf{Confidence}\tablenotemark{b}} 
}
\tablenotetext{a}{The luminosity limit of the sample.  The fit applies only to sources brighter than this limit.}
\tablenotetext{b}{The fraction of Monte Carlo tests with fits statistically worse than the fit to our sample.}
\startdata
Disk & 4.0$\times 10^{36}$  & 0.9$\pm$0.1 & 0.29\\
Disk & 1.3$\times 10^{37}$  & 1.4$\pm$0.2 & 0.93\\
GCs & 1.3$\times 10^{37}$  & 0.84$\pm$0.03 & 1.00\\
\enddata
\label{plfits}
\end{deluxetable}

\begin{deluxetable}{lcccc}
\tablewidth{6.5in}
\tablecaption{Peak fluxes and quiescent upper limits for transient candidates.}
\tableheadfrac{0.01}
\tablehead{
\colhead{{OBJ\tablenotemark{a}}} &
\colhead{{{Peak\tablenotemark{b}}} (10$^{36} erg\ s^{-1})$} &
\colhead{{Active Epochs}} &
\colhead{{{Q\tablenotemark{c}}}\ (10$^{36} erg\ s^{-1}$) {{(epochs)}}} &
\colhead{{{Peak/Q}}}
}
\tablenotetext{a}{$B$ labels denote transient events in the M31 bulge (within 7$'$ of the nucleus).}
\tablenotetext{b}{Peak luminosities are taken from the brightest observed epoch.}
\tablenotetext{c}{The 1$\sigma$ upper limit of the quiescent luminosity from combined quiescent epochs (number of quiescent epochs given in parenthesis).}
\tablenotetext{1}{Object detected at 10$^{37}$ \ergs in \citet{primini1993}, but undetected with upper limit of 5$\times$10$^{35}$ \ergs in \citet{kong2002}.}
\tablenotetext{2}{Known repeating transient SSS \citep{osborne2001}.}
\tablenotetext{3}{Known repeating transient SSS \citep{white1995}.}
\startdata
s1-79 &  152.89$\pm$13.81 & 1,3,4,5 & $\leq$2.33 (7) & $>$66\\%
s1-80 &  166.48$\pm$17.98 & 1,17 & $\leq$1.34 (11) & $>$124\\%
r3-125\tablenotemark{1} &  15.96$\pm$8.12 & 9,11 &$\leq$0.38 (11) & $>$42\\%
r3-46 ($B$) &  36.70$\pm$10.03 & 8,9 & $\leq$1.57 (9) & $>$23\\%
s1-82 &  16.45$\pm$8.02 & 1 & $\leq$1.83 (10) & $>$9\\%
r2-29 ($B$) &  38.78$\pm$9.76 & 7 & $\leq$0.43 (12) & $>$90\\%
r2-28 ($B$) &  15.89$\pm$2.87 & 2,3 & $\leq$1.60 (11) & $>$10\\%
s1-1 & 27.50$\pm$9.75 & 14,15,16 &  $\leq$2.44 (9) & $>$11\\%
r1-34 ($B$) &  42.98$\pm$1.31 & 14,15 & $\leq$2.10 (10) & $>$20\\%
s1-85 &  148.20$\pm$16.57 & 9,10 & $\leq$1.80 (9) & $>$82\\%
r2-69 ($B$) & 29.15$\pm$8.87 & 10 & $\leq$0.66 (10) & $>$44\\%
r2-8 ($B$) & 23.83$\pm$8.19 & 9,10,11 & $\leq$0.18 (10) & $>$132\\%
r2-67 ($B$) & 340.86$\pm$3.68 & 14,15,16 & $\leq$0.81 (10) & $>$421\\%
r3-115 ($B$) & 31.86$\pm$10.23 & 16 & $\leq$1.40 (8) & $>$23\\%
r3-126\tablenotemark{2} & 57.55$\pm$11.62 & 6,7 & $\leq$0.11 (8) & $>$52\\%
n1-85\tablenotemark{3} & 61.75$\pm$12.35 & 14 & $\leq$2.02 (11) & $>$34\\%
n1-59 & 33.10$\pm$13.88 & 10 & $\leq$1.16 (10) & $>$29\\%
\enddata
\label{tlist}
\end{deluxetable}

\oddsidemargin 0.0in

\clearpage

\begin{deluxetable}{cccc}
\tablewidth{5.0in}
\tablecaption{Optical magnitudes and
upper limits for sources contained within the regions surveyed by the
LGS.}  
\tableheadfrac{0.01} 
\tablehead{ 
\colhead{\bf{OBJ}} &
\colhead{\bf{Angular Separation ($''$)}} &
\colhead{\bf{B}} &
\colhead{\bf{V}}
}
\tablenotetext{a}{s1-75 is a known BL Lac candidate.}
\startdata
s1-74 &  $<$2   &    &  $<$14.7\\
s1-75\tablenotemark{a} &  0.48   &  21.46$\pm$0.11  &  20.45$\pm$0.11\\
s1-76 &    &   &  $>$20.7\\
s1-78 &  0.85   &  20.67$\pm$0.11  &  20.11$\pm$0.11\\
s2-63 &    &   &  $>$20.9\\
s1-45 &  $<$2   &   &  $<$14.7\\
s1-34 &    &   &  $>$21.3\\
s1-64 &  1.6   &  21.12$\pm$0.14  &  21.23$\pm$0.14\\
r2-24 &    &   &  $>$19.5\\
r3-31 &    &   &  $>$20.35\\
r1-30 &    &   &  $>$19.5\\
r2-18 &    &   &  $>$19.5\\
r2-16 &    &   &  $>$19.5\\
r3-28 &  1.3   &  21.96$\pm$0.16  &  22.39$\pm$0.35\\
r1-2 &    &   &  $>$19.5\\
r3-25 &    &   &  $>$20.75\\
r2-14 &    &   &  $>$19.5\\
r2-13 &    &   &  $>$19.5\\
r3-69 &    &   &  $>$20.75\\
r2-10 &    &   &  $>$19.5\\
r3-23 &    &   &  $>$20.75\\
r2-8 &  1.21   &  21.80$\pm$0.16  &  20.98$\pm$0.16\\
n1-76 &    &   &  $>$20.9\\
r2-67 &  0.54   &  22.34$\pm$0.27  &  21.89$\pm$0.11\\
r3-67 &    &   &  $>$20.1\\
r3-115 & 1.8   & 23.30$\pm$0.41  &  21.96$\pm$0.10\\
r3-16 &  0.73   &  19.07$\pm$0.14  &   18.84$\pm$0.13\\
r3-13 &  1.04   &  21.30$\pm$0.17  &   21.28$\pm$0.15\\
r3-9 &    &   &  $>$20.1\\
r3-8 &    &   &  $>$20.1\\
r3-126 &    &   &  $>$20.1\\
r3-7 &  1.51   &  20.64$\pm$0.14  &  20.56$\pm$0.14\\
r3-63 &    &   &  $>$20.55\\
n1-77 &    &   &  $>$20.45\\
n1-79 &    &   &  $>$20.7\\
n1-80 &    &   &  $>$21.4\\
n1-81 &  0.49   &  21.09$\pm$0.11  &  20.58$\pm$0.11\\
n1-82 &  1.32   &  15.51$\pm$0.10  &  14.95$\pm$0.10\\
n1-83 &    &   &  $>$21.7\\
n1-84 &    &   &  $>$21.6\\
n1-61 &    &   &  $>$21.5\\
n1-59 &  1.50   &  17.88$\pm$0.10  &  17.15$\pm$0.10\\
\enddata
\label{lgslist}
\end{deluxetable}

\begin{deluxetable}{ccccc}
\tablewidth{0pt}
\tablecaption{Results of HST/WFPC2 and LGS observations for r2-67 (CXOM31 J004305.6+411703)}
\tableheadfrac{0.01}
\tablehead{
\colhead{\bf{Date}} &
\colhead{\bf{Filter}} &
\colhead{\bf{Exposure (s)}} &
\colhead{\bf{magnitude}} 
}
\startdata
2000-10-06 & $B$ & 300 & 22.34$\pm$0.27\\
2001-09-18 & $V$ & 300 & 21.89$\pm$0.11\\
2001-11-05 & F336W (U) & 4000 & 22.32$\pm$0.15\\
2001-12-02 & F336W (U)  & 4000 & 22.32$\pm$0.11\\
2002-02-04 & F336W (U)  & 4000 & $\geq$22.8\\
\enddata
\label{r267obs}
\end{deluxetable} 

\begin{deluxetable}{ccccc}
\tablewidth{0pt}
\tablecaption{Results of HST/WFPC2 and LGS observations for r3-16 (CXOM31 J004309.8+411901)}
\tableheadfrac{0.01}
\tablehead{
\colhead{\bf{Date}} &
\colhead{\bf{Filter}} &
\colhead{\bf{Exposure (s)}} &
\colhead{\bf{magnitude}} 
}
\startdata
2000-10-06 & $B$ & 300 & 19.07$\pm$0.14\\
2001-08-27 & F336W (U) & 4000 & 20.82$\pm$0.06\\
2001-09-18 & $V$ & 300 & 18.84$\pm$0.13\\
2001-12-02 & F336W (U) & 4000 & 20.82$\pm$0.06\\
2002-01-08 & F336W (U)  & 4000 & 21.11$\pm$0.02\\
\enddata
\label{r316obs}
\end{deluxetable}

\clearpage

\clearpage
\vspace{-1.5cm}

\begin{figure}
\centerline{\psfig{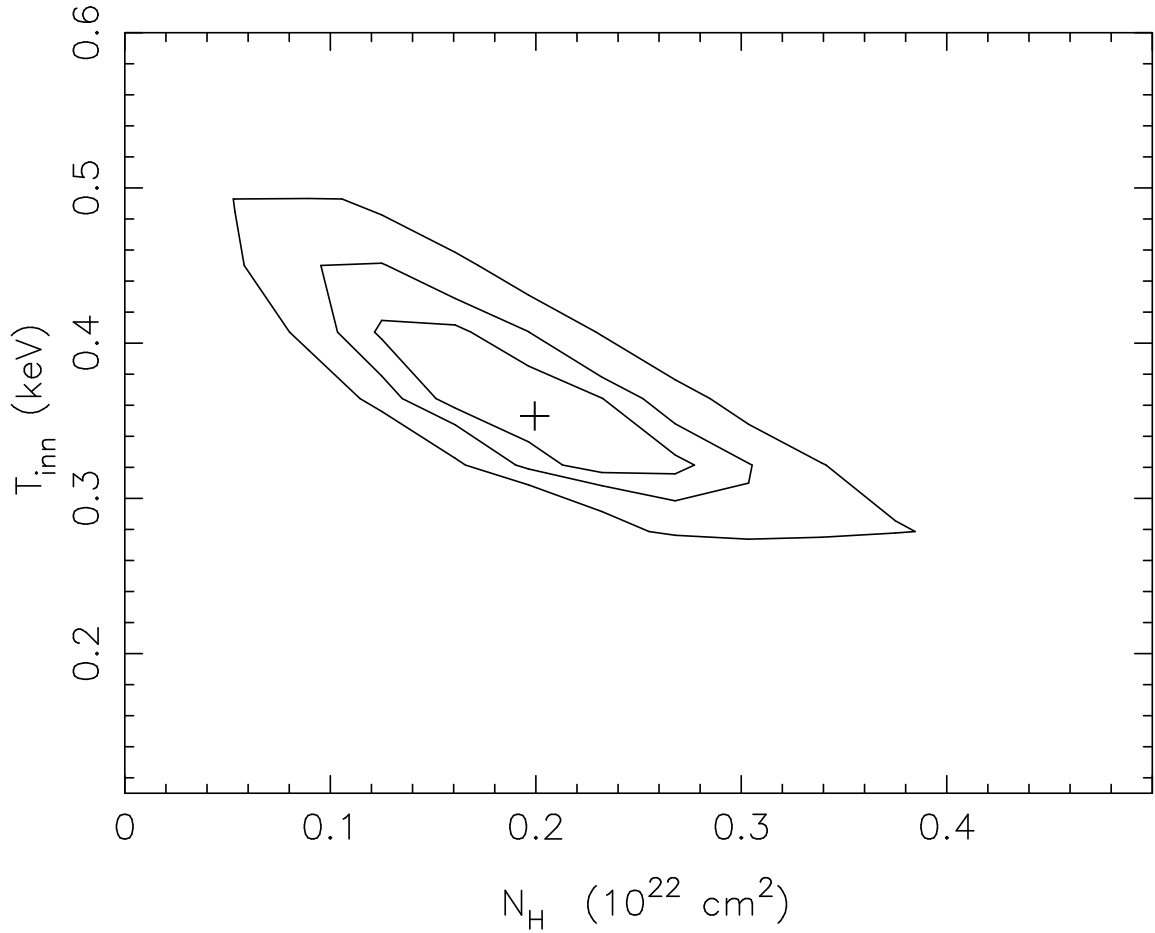}}
\caption{Spectral parameter confidence contours for the 2001 Nov. 19
observations of r2-67 (OBSID~1585) as determined via ISIS using the
full pile-up model.  The measured temperature of the inner edge of the
accretion disk is $0.35\pm0.05$ keV.  The absorption value of $\rm N_H
= 2\pm1 \times 10^{21}$ cm$^2$ corresponds to an expected $A_U =
1.7\pm0.9$.}
\label{obs1585}
\end{figure}

\clearpage

\begin{figure}
\centerline{\psfig{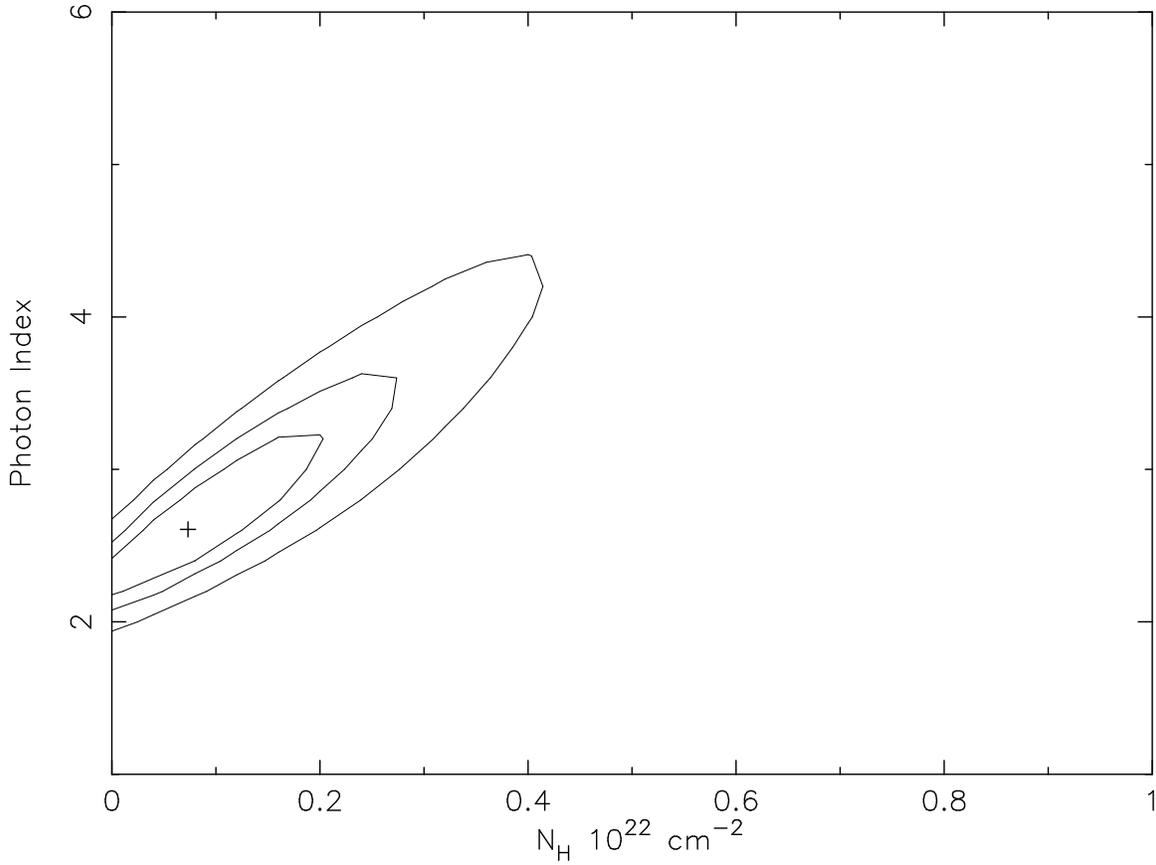}}
\caption{Spectral parameter confidence contours for the 2002 Jan. 8
observations of r2-67 (OBSID~2897).  The $\rm N_H$ ($0.7\pm1.3 \times
10^{21}$ cm$^2$) partially overlaps at the 1$\sigma$ level with the
earlier observation.  Combining this $\rm N_H$ value and the
consistent value measured from OBSID 1585 yields our best estimate of
the absorption toward r2-67: $\rm N_H = 1.5\pm0.8 \times 10^{21}$
cm$^2$, or $A_U = 1.3\pm0.7$}
\label{obs2897}
\end{figure}

\newpage
\clearpage

\begin{center}
{\bf APPENDIX}
\end{center}

In fitting the ACIS-I spectrum of r2-67 from OBSID 1585 using a pileup
model, we limit ourselves to the 0.3--7.0~keV range in order to
exclude background counts, and we include the `afterglow' events,
which account for 92 of the 866 total good events.  We do not find a
clear change in ACIS grades in radial averaged profiles but note that
this is likely due to our much lower counting rate than that seen from
GX~13+1 \citep{smith2002}.

Given the high luminosity of the source during the outburst, we should
expect the source to show a thermal spectrum in the shape of a disk
blackbody \citep{mcclintock2004, mitsuda1984} rather than a power-law.
However, fits to both shapes while neglecting pileup may yield useful
comparisons to fits including pileup.  Fits to a disk blackbody yield
$\chi^2_{\nu} = 1.8$ (probability $\sim 10^{-3}$), $T_{inn} =
0.95$~keV, $\rm N_H = 2.8 \times 10^{21}$ cm$^2$.  Fits to a power-law
find $\chi^2_{\nu} = 1.8$, $\alpha = 2.5$, $\rm N_H = 2.5 \times
10^{21}$ cm$^2$.  Both these fits show a clear excess of counts
between 2 and 3 keV, which is indicative of pileup \citep{nowak2002a}.
A simple blackbody is a very poor representation of the data, yielding
at best $\chi^2_{\nu} = 18$.

Because the pileup model was first developed within ISIS and later
incorporated into Sherpa and XSPEC, we used ISIS to determine the
pileup parameters.  Allowing the pileup and spectral model parameters
to vary (but freezing the PSF fraction to 0.95), ISIS quickly
converges to a model with 30\% pileup, a grade migration parameter
$\alpha_G = 0.99$.  With the pileup parameters frozen at the best
values, ISIS finds spectral parameters of $T_{inn} = 0.35\pm 0.05
$~keV, $N = 11^{+20}_{-8} (R_{inn}/km)^2 (10/d)^2 cos (\theta)$, and
$\rm N_H = 2.0\pm 1.0 \times 10^{21}$cm$^{-2}$ with $\chi^2_{\nu}$ =
1.2 (probability=0.2), where $T_{inn}$ is the temperature of the inner
edge of the accretion disk, $N$ is the normalization parameter, $d$ is
the distance to the source, and $\theta$ is the inclination angle of
the disk.  The $\chi^2$ contours are shown in Figure~\ref{obs1585}.
The observed flux is 1.3 $\times$ 10$^{-12}$ \ergcm2s (0.3-7.0 keV),
and the modeled emitted luminosity is $1.9 \times 10^{38}$ \ergs
(0.3-7.0 keV).

We note that when pile-up is included, we are not able to rule-out a
power-law model solely on the basis of $\chi^2$.  A power-law fit
gives a slightly lower $\alpha_G \sim 0.8$, a slope $\alpha = 4.4 \pm
0.5$, a much higher ${\rm N_H} = 5\pm 1 \times 10^{21}$ cm$^2$, but
has an acceptable $\chi^2/\nu = 1.2$.  While this model has the same
observed flux, the steep slope and higher ${\rm N_H}$ predicts a
higher emitted luminosity of $2.1 \times 10^{39}$\ergs .  We discount
the power-law model for three reasons.  First, it is not the thermal
form expected at this high flux.  Second, the ${\rm N_H}$ is
inconsistent with that measured for the same source many months later.
Third, this high ${\rm N_H}$ predicts $A_U = 4.3$ and the unreasonable
high absolute magnitude of M$_U=-6.5$ for the HST counterpart.

Fits with XSPEC produced similar results, but were more sensitive to
the initial guesses for the spectral parameters and were insensitive
to the value of $\alpha_G$.  We therefore fixed $\alpha_G$ to 0.99 and
and the PSF fraction to 0.95.  The best fit disk black body finds
$T_{inn} = 0.36 ^{+0.07}_{-0.04}$, ${\rm N_H} = 2 \pm 0.7 \times
10^{21}$ cm$^2$, both of which overlap with the ISIS determined
values.

Fits with Sherpa and the pileup model produced spectral parameters
consistent with those determined by ISIS.  Sherpa found a pileup grade
migration parameter $\alpha_G = 1$ and a pileup fraction of 30\%.  The
best fit disk spectral parameters were $T_{inn} = 0.36 \pm 0.02 $~keV
and $N = 9 \pm 3 \  (R_{inn}/km)^2 (10/d)^2 cos (\theta)$ $\rm N_H =
1.9\pm 0.4 \times 10^{21}$cm$^{-2}$ , with $\chi^2_{\nu}$ = 1.3
(probability=0.1).

One check on the results of the pileup model fits which has been used
previously is to extract the spectrum of the source from only with
wings of the PSF, where the counting rate is low enough that pileup
can be ignored (\citealp{swartz2003} on M81).  While this can 
give a qualitative measure, the absolute value of the
parameters determined this way must be treated with care 
because hard photons are preferentially
scattered into the wings of the PSF and the current effective area
tools (i.e.  mkarf) do not take this into account
(i.e. \citealp{smith2002}).

We tried ignoring pileup and removing the central pixel only and using
the surrounding 8 pixels (leaving 450 counts), and removing the
central 9 pixels and using the surrounding 40 (leaving 107 counts).
The first method yields $\chi^2_{\nu} = 1.3$ (prob = 14\%) and shows a
significant excess ($\sim$45\% above model) of counts between 2 and 3
keV. The best fit values are T$_{inn} = 0.9 \pm 0.1$~keV and $\rm N_H
< 0.6 \times 10^{21}$\cm-2 . The second method yields $\chi^2_{\nu} =
2.0$ (prob = 5\%) and similar excess between 2 and 3 keV.  The best
fit values are T$_{inn}= 0.74 \pm 0.12$~keV and $\rm N_H < 1 \times
10^{21}$\cm-2 .

Removing the central pixel(s) appears to decrease the amount of
pileup, as evidenced by the slight decrease in the excess (above
model) of counts between 2 and 3 keV.  However, the fitted
temperatures are much higher than and completely inconsistent with 
the temperature found using all the data and accounting for pileup.  
It is unclear to us if these erroneously high temperatures are due
to the scattering of hard photons into the wings or due to ignoring 
the effects of pileup, since both have the effect of hardening the
fitted spectrum. 

\end{document}